\documentclass[lettersize,journal]{IEEEtran}
\usepackage{amsmath,amsfonts}
\usepackage{algorithmic}
\usepackage{algorithm}
\usepackage{array}
\usepackage[caption=false,font=normalsize,labelfont=sf,textfont=sf]{subfig}
\usepackage{textcomp}
\usepackage{stfloats}
\usepackage{url}
\usepackage{verbatim}
\usepackage{graphicx}
\usepackage{cite}
\usepackage{booktabs}
\usepackage{multirow}
\usepackage[table]{xcolor}
\usepackage{amssymb}
\def\BibTeX{{\rm B\kern-.05em{\sc i\kern-.025em b}\kern-.08em
    T\kern-.1667em\lower.7ex\hbox{E}\kern-.125emX}}
\usepackage{balance}

\usepackage{graphicx} 
\usepackage{hyperref} 
\usepackage[all]{hypcap}
\usepackage{cleveref}

\usepackage{xcolor}
\usepackage{soul}      
\soulregister\cite7
\soulregister\Cite7
\soulregister\parencite7
\soulregister\textcite7
\soulregister\cref7
\soulregister\Cref7
\soulregister\ref7
\soulregister\eqref7

\soulregister\cred7

\usepackage{ifthen}

\newif\ifrevmode
\revmodetrue          

\newcommand{\orcidicon}{\includegraphics[scale=0.6]{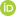}} 

\newcommand{\orcidlink}[1]{\href{https://orcid.org/#1}{\orcidicon}}

\begin{document}

\title{Modality-Invariant Coarse-to-Fine Retinal Image Registration}
\author{Bo Wen \raisebox{1.5pt}{\orcidlink{0009-0001-5747-4739}}, \IEEEmembership{Graduate Student Member, IEEE}, Nehal Nailesh Mehta \raisebox{1.5pt}{\orcidlink{0000-0003-4653-2021}}, Melanie Tran \raisebox{1.5pt}{\orcidlink{0009-0001-7533-2726}}, \\ Dirk-Uwe Bartsch \raisebox{1.5pt}{\orcidlink{0000-0003-0955-8708}}, William Freeman \raisebox{1.5pt}{\orcidlink{0000-0001-9979-2500}}, Truong Nguyen \raisebox{1.5pt}{\orcidlink{0000-0002-5022-063X}}, \IEEEmembership{Fellow, IEEE}} 

\definecolor{lightblue}{rgb}{0.8,0.9,1}
\definecolor{lightgreen}{rgb}{0.8,1,0.9}
\definecolor{lightred}{rgb}{0.9,0.6,0.6}



\maketitle

\crefname{table}{Table}{Tables}
\crefname{figure}{Fig.}{Figs.}

\begin{abstract}
Retinal image registration is essential for ophthalmic diagnosis, longitudinal disease monitoring, and multimodal retinal image analysis. Existing retinal registration methods are typically modality-dependent: they are designed or optimized either for a single imaging modality in mono-modal registration or for a fixed pair of modalities in cross-modal registration. This limits their flexibility and applicability in practical scenarios involving diverse retinal imaging modalities and different combinations of them. In this work, we propose a generalizable two-stage, modality-invariant framework for retinal image registration. First, we introduce a sparse feature-matching model driven by a universal retinal vessel segmentation to achieve robust coarse global alignment across modalities. Second, we develop a modality-invariant optical flow estimation network, termed MI-RAFT, to refine the alignment through dense local registration. Extensive experiments demonstrate that the proposed method can handle diverse combinations of commonly used retinal imaging modalities, exhibiting strong modality invariance while outperforming state-of-the-art modality-dependent registration methods. \footnote{This paper is a submission to IEEE Transactions on Image Processing. The code of this work will be available at the time of publication.} \footnote{This work was supported by the National Institute of Health, USA under Grant R01 EY033847-02.}
\end{abstract}



\section{Background and Introduction}
\IEEEPARstart{R}{etinal} image registration is an important task in medical image processing. In particular, accurate registration of cross-modality retinal images enables the co-localization of complementary multimodal information, allowing clinicians to identify and diagnose retinal diseases more accurately and efficiently \cite{Intro-medical_background}. It also has numerous other applications such as retinal eye tracking \cite{Application-eye_tracking}, retina-based biometric authentication \cite{Registration-SuperRetina} and retinally stabilized stimulus delivery \cite{Application-stability}. However, to the best of our knowledge, existing methods for both single-modality and cross-modality retinal image registration are developed and validated in a modality-dependent manner. Specifically, cross-modality methods typically require fine-tuning for, and are applicable only to, a fixed pair of modalities. In contrast, we propose a more practical, modality-invariant solution that robustly and accurately registers diverse combinations of modalities, including both single-modality and cross-modality image pairs. This capability reflects the real-world need for a generalizable solution to retinal image registration.

Moreover, global registration based on sparse features alone is insufficient to accurately align all regions of an image. To achieve precise retinal image alignment, state-of-the-art methods commonly adopt a two-stage, coarse-to-fine pipeline \cite{Registration-Zhang-TIP, Intro-coarse2fine_background, Registration-Ding-coarse2fine, Registration-Liu-GAMorph}, in which images are first globally aligned by estimating sparse feature correspondences and then finely aligned using optical flow. In this work, we propose a complete two-stage framework that achieves modality invariance in both coarse and fine alignment.

\begin{figure}[ht]
\centering
\includegraphics[width=0.49\textwidth]{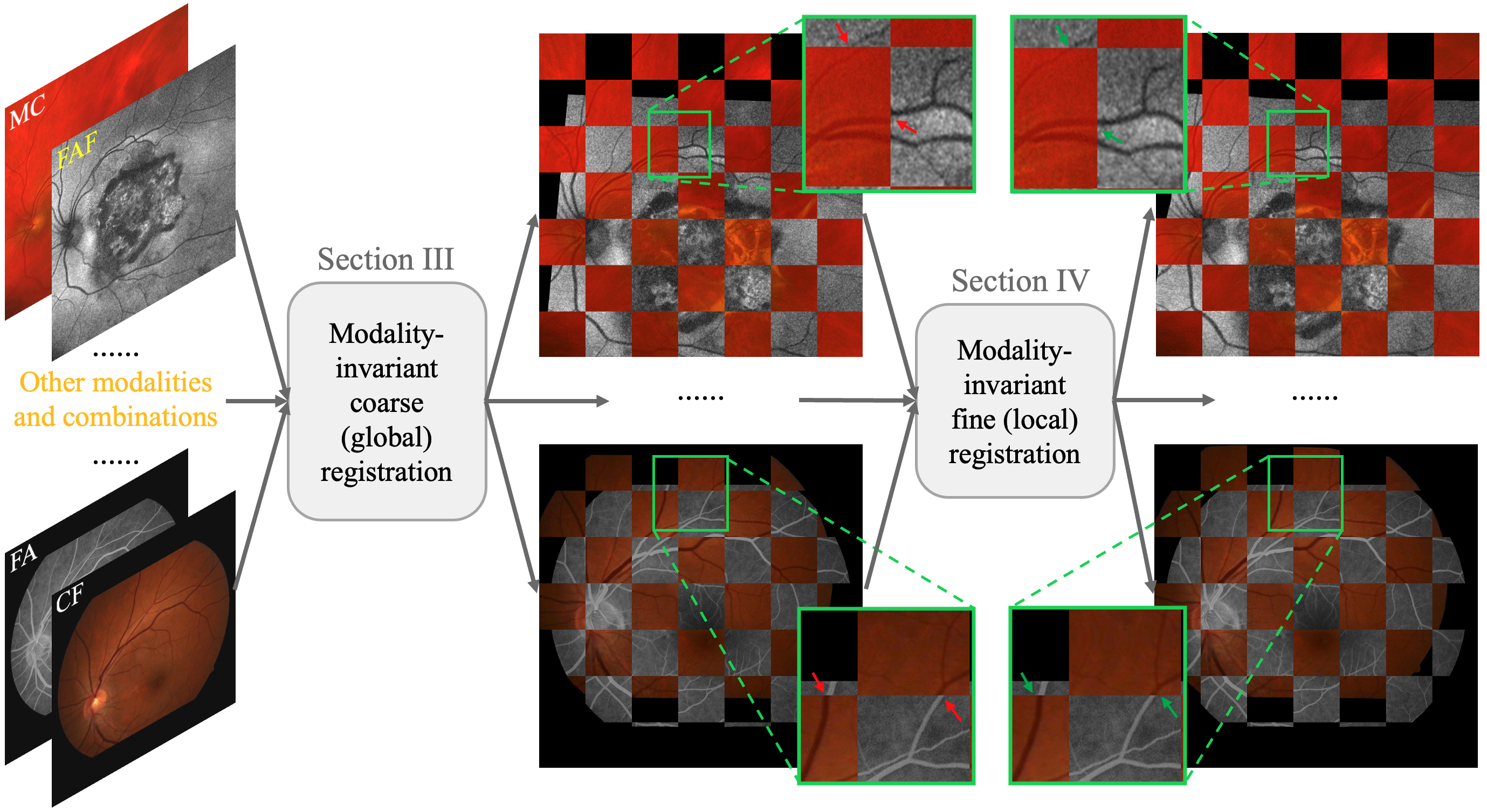}
\vspace{-0.3cm}
\caption{Modality-invariant coarse (global, linear)-to-fine (local, deformable) retinal image registration (best zoom in to view and viewed in color).}
\label{fig:intro}
\vspace{-0.3cm}
\end{figure}

For coarse global alignment, we propose using a Universal Retinal Vessel Segmentation Model \cite{Segmentation_Wen_TIP} as a modality-invariant domain adapter that transforms retinal images from different modalities into a unified vessel segmentation domain. Consequently, a sparse feature detection and outlier rejection model trained on a single modality pair can generalize robustly to other modalities and modality combinations. Although the proposed framework is designed to be invariant to imaging modality, its current formulation is not intended to address extreme differences in field of view (FOV). In particular, registration between a 135$^\circ$ ultra-widefield \cite{Segmentation_Wen_TIP, Registration-Kalaw-Eye} image and a conventional 30$^\circ$ fundus image remains challenging because of the substantial scale difference and limited anatomical overlap. We therefore define the scope of this work as modality-invariant registration among conventional 30--45$^\circ$ en face retinal images with comparable FOVs. This setting nevertheless encompasses diverse mono-modal and cross-modal combinations among commonly used retinal imaging modalities. Extending the framework to jointly address variations in both modality and FOV represents an important direction for future work toward fully universal retinal image registration.

For fine local alignment, although previous studies have explored cross-modal optical flow and multimodal dense registration, a single model capable of estimating dense registration fields across arbitrary retinal modality combinations without modality-pair-specific adaptation remains largely unexplored. To address this gap, we propose a novel Modality-Invariant Recurrent All-Pairs Field Transforms (MI-RAFT) model for estimating optical flow between multimodal retinal images. Previous studies have demonstrated that features extracted from diffusion models \cite{Others-stable_diffusion} or large pretrained vision backbones \cite{Others-DINOV3} can serve as emergent correspondence descriptors \cite{Correspondence-DIFT-NeurIPS, Correspondence-Zhang-NeurIPS, Correspondence-Amir-ECCVW} and exhibit strong cross-modality generalization \cite{Correspondence_MIFNet_TIP}. However, these features have primarily been used for sparse correspondence matching rather than incorporated into an optical flow architecture for iterative dense field estimation. MI-RAFT introduces an effective design for leveraging such features in dense optical flow estimation. Specifically, it learns to refine and fuse flow-optimized features with modality-invariant features and uses the resulting representation for iterative flow estimation and refinement. We further present a synthetic-pair generation process that enables MI-RAFT to be trained using unpaired retinal images, eliminating the need to collect a large paired dataset while providing direct ground-truth flow supervision. Besides, we propose a correlation lookup feature alignment training strategy which further enhances the modality-invariance of the model and quality of the predicted flow. Notably, despite being trained entirely on synthetically generated pairs, MI-RAFT outperforms state-of-the-art modality-dependent methods fine-tuned on real image pairs from the corresponding modality combinations.

We validate the proposed methods on diverse mono-modal and cross-modal retinal image datasets. We further collect a new dataset of paired fundus autofluorescence (FAF) and multicolor scanning laser ophthalmoscopy (MC) retinal images to address the absence of registration dataset in these two widely-used modalities. Extensive experiments demonstrate the superior performance of our method over state-of-the-art methods for retinal image registration, modality-invariant correspondence estimation, and cross-modal optical flow estimation. The main contributions of this work are summarized as follows:
\begin{itemize}
\item We formulate the task of modality-invariant retinal image registration and, to the best of our knowledge, propose the first modality-invariant retinal image registration framework that addresses both global (coarse) and local (fine) alignment. This formulation reflects the practical need for generalizable and comprehensive retinal image registration solutions.
\item We demonstrate that a recent universal retinal vessel segmentation model can serve as an effective modality-invariant domain adapter. When integrated into a standard sparse-feature-based registration framework, it enables state-of-the-art performance in modality-invariant global retinal image registration.
\item We propose a novel modality-invariant optical flow estimation model, termed Modality-Invariant(MI)-RAFT, for fine local retinal image registration. Its key technical contributions include a new architectural design, a synthetic dual-training-pair generation scheme jointed with a correlation-lookup feature-alignment strategy.
\item Extensive experiments on a wide variety of mono-modal and cross-modal retinal image registration datasets demonstrate the robustness of our framework and its superior performance over state-of-the-art baselines from retinal image registration and related correspondence and optical flow literature.
\end{itemize}

\section{Related Work}
\label{sec:literature}

\subsection{Retinal Image Registration}
Retinal image registration geometrically aligns retinal images acquired at different times, from different viewpoints, or using different imaging devices or modalities. Deep learning-based approaches have become increasingly prevalent in this field in recent years. Mahapatra et al. \cite{Registration-Mahapatra} introduced an unsupervised deformable registration approach based on generative adversarial networks, which directly synthesizes the registered image and estimates the corresponding deformation field. Luo et al. \cite{Registration-Luo} proposed a CNN to estimate affine transformation matrices between indocyanine green angiography (ICGA) and multicolor (MC) images. Arikan et al. \cite{Registration-Arikan} developed a multimodal retinal image registration method that uses two CNNs to extract vessel segmentations and identify vascular bifurcations for alignment. Tian et al. \cite{Registration-Tian} proposed an unsupervised deformable registration network that uses image gradients from both input images as alignment cues during training. Lee et al. \cite{Registration-Lee} presented a feature-filtering CNN designed to detect and exclude unreliable step-pattern features in multimodal retinal image registration. Wang et al. \cite{Registration-Wang-TIP} proposed a segmentation-based registration framework that establishes correspondences between keypoints detected on vessel segmentation maps. Zhang et al. \cite{Registration-Zhang-TIP} proposed a two-stage registration framework that further refines an initial global alignment using an optical-flow-based fine-alignment algorithm. In subsequent work, Zhang et al. \cite{Registration-Wen-ICIP} introduced a distortion-correction method based on 3D eyeball-surface optimization for aligning ultra-widefield and conventional retinal images. Liu et al. \cite{Registration-SuperRetina} proposed SuperRetina, a semi-supervised keypoint detection and description method that uses progressive keypoint expansion to enrich incomplete keypoint annotations. They further introduced an improved keypoint-based triplet loss for descriptor learning. Wang et al. \cite{Registration-SuperJunction} proposed SuperJunction, a learning-based keypoint detection and description framework that incorporates vessel detection as a regularization task and employs constrained negative sampling to improve descriptor learning, together with a hybrid matching strategy for handling nonlinear deformations across retinal views. Liu et al. \cite{Registration-Liu-GAMorph} proposed HybridRetina, a progressive coarse-to-fine framework that combines keypoint-based global homography estimation with a geometry-aware deformation network, termed GAMorph, guided by multilevel pixel relationships and edge attention for local retinal image alignment. Most recently, Liang et al. \cite{Registration-Liang-EyeKey} proposed EyeKey, a self-supervised keypoint detection and description network that leverages local feature saliency and a detect-while-describing design for efficient mono-modal and multimodal retinal image global registration.

\subsection{Cross-Modal Optical Flow Estimation}
Cross-modal optical flow estimation aims to establish dense pixelwise correspondences between images from different modalities, for which the conventional brightness-constancy assumption generally does not hold. Zhou et al. \cite{Registration-CrossRAFT} proposed CrossRAFT, which transfers the prior knowledge of mono-modal optical flow networks to diverse cross-modal scenarios through a self-supervised modality-promotion framework and introduces a cross-modal adapter into RAFT to improve feature compatibility across modalities. Zhai et al. \cite{Registration-MCMRAFT} subsequently proposed MCM-RAFT, which employs a modality compensation module to adaptively exchange complementary information between cross-modal features and a feature-alignment loss to reduce their representational discrepancy. Beyond general-purpose cross-modal flow estimation, Zhang et al. \cite{Registration-Zhang-JSTARS} developed OSFlowNet-Ft for dense optical--SAR image registration, incorporating dilated feature concatenation to enhance pixelwise feature discrimination, synthetic topographic deformations for supervised training, and self-supervised instance-specific fine-tuning based on blockwise matching. Sun et al. \cite{Registration-Sun-GDROS} proposed GDROS for optical--SAR registration under large geometric transformations, which constructs and iteratively refines multi-scale 4D correlation volumes from hybrid CNN--Transformer features and constrains the resulting dense flow using an affine transformation estimated through least-squares regression. More recently, Zhang et al. \cite{Registration-Zhang-DCFlow} introduced DCFlow, which decouples modality transfer from flow estimation and jointly coordinates them through cross-modal consistency; while requiring no ground-truth flow for real cross-modal pairs, it trains the flow estimator with geometry-derived synthetic flow labels generated from individual images. Liu et al. \cite{Registration-Liu-CRFT} proposed CRFT, a coarse-to-fine transformer framework that learns cross-modality feature flow and recurrently refines the estimated field using discrepancy-guided attention and spatial geometric transformations.

\subsection{Modality-Invariant Image Correspondence}
Most learning-based multimodal image matching methods are developed for predefined modality pairs and therefore generalize poorly to unseen modalities or modality combinations. To overcome this limitation, recent studies have shifted from modality-pair-specific matching toward learning correspondences that generalize across arbitrary modalities. Liu et al. \cite{Correspondence_MIFNet_TIP} proposed MIFNet, which refines latent Stable Diffusion features and fuses them with conventional keypoint descriptors to learn modality-invariant representations from only mono-modal training images. Ren et al. \cite{Correspondence-Ren-MINIMA} proposed MINIMA, which instead addresses modality generalization through data scaling: a generative data engine converts labeled RGB image pairs into a large synthetic multimodal dataset while preserving their correspondence labels, enabling sparse, semi-dense, and dense matching pipelines to be trained once and applied across diverse cross-modal scenarios.

\section{Universal Retinal Vessel Segmentation-driven Modality-Invariant Coarse Alignment}
\label{sec:coarse}
Despite recent advances in modality-invariant image correspondence, we observe that existing methods, even when fine-tuned on retinal images, often fail to establish correspondences with the accuracy and reliability required for precise retinal image alignment. An effective alternative is to use vessel segmentation maps as a modality-independent intermediate representation, thereby reducing the appearance gap among retinal imaging modalities. Previous studies have demonstrated that detecting and matching vascular features in this shared representation can achieve more accurate cross-modal retinal image registration than directly extracting and matching features from the original images \cite{Registration-Wang-ICASSP, Registration-Wang-TIP, Registration-Zhang-TIP}.

Nevertheless, existing vessel-segmentation-based registration methods are generally constrained by the modality dependence of their underlying segmentation models, limiting their ability to generalize across arbitrary retinal modalities. Recently, the Universal Retinal Vessel Segmentation Model (URVSM) \cite{Segmentation_Wen_TIP} achieved modality-invariant vessel segmentation by combining image translation with topology-aware feature learning for domain adaptation. In this work, we further investigate its utility for retinal image registration and demonstrate that URVSM can serve as a reliable modality-invariant domain adapter, transforming retinal images from diverse modalities into a shared vascular representation suitable for accurate correspondence estimation.

\begin{figure*}[ht]
\centering
\includegraphics[width=1\textwidth]{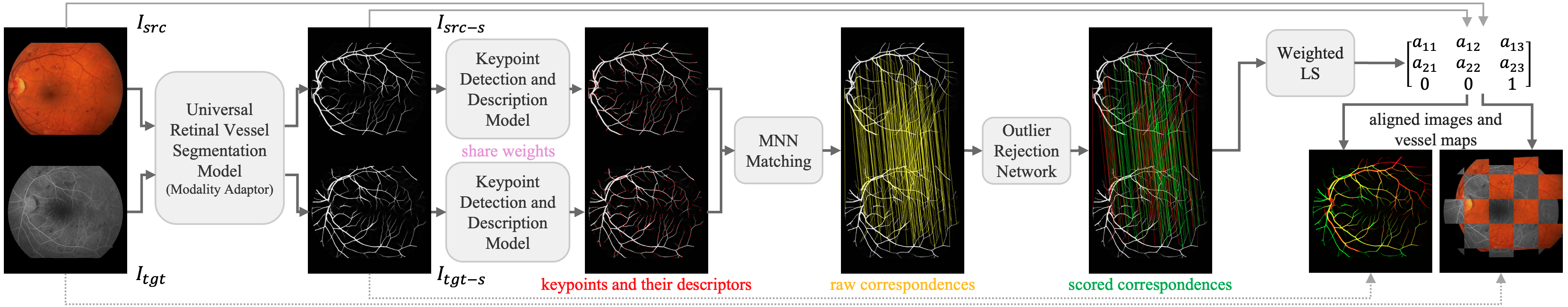}
\vspace{-0.3cm}
\caption{Pipeline of the proposed modality-invariant coarse/global retinal image registration.}
\vspace{-0.3cm}
\label{fig:coarse_pipeline}
\end{figure*}

The pipeline of the coarse-alignment algorithm is illustrated in Fig.~\ref{fig:coarse_pipeline}. Given two retinal images (the goal is to warp and align a source image $I_{src}$ to a target image $I_{tgt}$) from arbitrary modalities, we first process both images using URVSM to obtain modality-independent vessel segmentation maps. The subsequent registration procedure follows established sparse-feature-based retinal image registration frameworks; therefore, we describe only its essential components and refer readers to the original papers for implementation details. The two vessel maps are processed by a shared keypoint detection and description network, which produces a keypoint probability map and a dense descriptor map for each image. We initialize the network using the pretrained weights from Wang et al. \cite{Registration-Wang-TIP}, whose model adopts SuperPoint \cite{Registration-SuperPoint} as its backbone. We freeze the shared encoder and keypoint detector head and fine-tune only the descriptor head on URVSM-generated vessel maps using the same paired color fundus (CF) and infrared reflectance (IR) training dataset as in \cite{Registration-Wang-TIP}. Following SuperRetina \cite{Registration-SuperRetina}, descriptor learning is supervised using the modified triplet loss:

\begin{equation}
    \mathcal{L}_{\mathrm{desc}}
    =
    \sum_{i \in \mathcal{K}}
    \max\left(
        0,\,
        m + \phi_{\mathrm{pos}}
        - \frac{1}{2}
        \left(
            \phi_{\mathrm{neg\text{-}rand}}
            + \phi_{\mathrm{neg\text{-}hard}}
        \right)
    \right)
\end{equation}
where $\mathcal{K}$ denotes the set of sampled keypoints and $m$ is the triplet-loss margin. We refer readers to \cite{Registration-SuperRetina} for the detailed formulation of this loss and to \cite{Registration-Wang-TIP} for the procedure used to obtain the weakly supervised ground-truth transformation matrices and determine positive and negative keypoint pairs according to these transformations.

Raw correspondences are established by mutual nearest-neighbor (MNN) matching between the keypoint descriptors extracted from $I_{\mathrm{src}}$ and $I_{\mathrm{tgt}}$. Specifically, a source keypoint $p_m$ is matched to a target keypoint $p'_n$ if and only if their descriptors $d_m$ and $d'_n$ satisfy
\begin{equation}
    n = \underset{j}{\arg\min}\,\lVert d_m-d'_j\rVert_2
    \quad \land \quad
    m = \underset{i}{\arg\min}\,\lVert d_i-d'_n\rVert_2
\end{equation}
The coordinates of the resulting correspondence pairs are subsequently fed into the pretrained outlier-rejection network of Wang et al. \cite{Registration-Wang-TIP}, which adopts the architecture proposed in \cite{Correspondence_CLe_CVPR}, to predict a weight $\in[0, 1]$ for each correspondence. We use this network without further fine-tuning because it operates solely on keypoint coordinates rather than image appearance or descriptors. Moreover, our keypoint locations are produced by the frozen detector from \cite{Registration-Wang-TIP}, which retains its original SuperPoint weights \cite{Registration-SuperPoint} in both their framework and ours. Consequently, the input distribution of the outlier-rejection network remains largely unchanged. Despite requiring no additional adaptation, this pretrained network yields more reliable correspondence filtering than the alternative outlier-rejection/learned matching methods evaluated in our experiments.

Finally, the retained correspondences and their predicted weights are used to estimate an affine transformation matrix via weighted least squares, following \cite{Registration-Zhang-TIP}. The estimated transformation is then applied to warp either the source vessel map or the original source image into the target coordinate system, completing the coarse-alignment stage.

\section{MI-RAFT: A Modality-Invariant Optical Flow Estimation Model for Fine Deformable Retinal Image Alignment}
\label{sec:fine}

Following the modality-invariant coarse/global registration stage, we introduce our proposed method for fine/local retinal image alignment. Features extracted from pretrained generative models, such as Diffusion Models \cite{Correspondence-DIFT-NeurIPS}, and discriminative models, such as DINO \cite{Others-DINOV3}, have emerged as powerful image-correspondence descriptors with strong generalization capabilities \cite{Correspondence-Zhang-NeurIPS,Correspondence-Amir-ECCVW}. However, these pretrained features are often spatially too coarse to provide the pixel-level precision required for task-specific dense correspondence estimation. Recent work on sparse correspondence estimation has shown that refining DIFT \cite{Correspondence-DIFT-NeurIPS} features and fusing them with task-specific keypoint features can produce more reliable modality-invariant descriptors \cite{Correspondence_MIFNet_TIP}. Nevertheless, the refinement and fusion of pretrained vision features for modality-invariant dense optical flow estimation remain largely unexplored. To address this limitation, we propose Modality-Invariant Recurrent All-Pairs Field Transforms (MI-RAFT), an optical flow architecture that effectively refines, fuses, and leverages pretrained vision features to estimate modality-invariant dense correspondences for fine deformable retinal image alignment.

\subsection{General Model Design and Explanation}
\begin{figure*}[ht]
\centering
\includegraphics[width=1\textwidth]{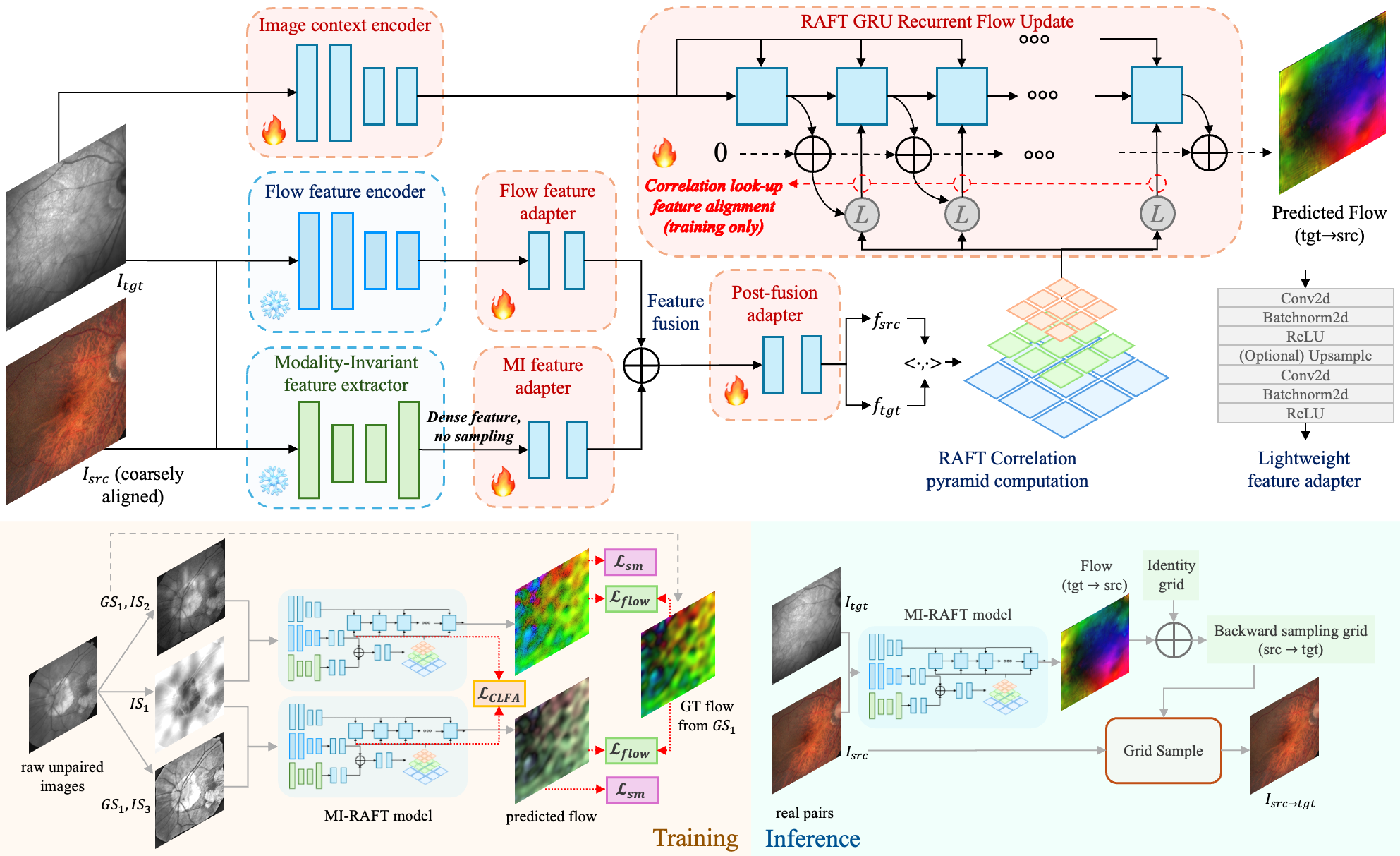}
\vspace{-0.1cm}
\caption{Proposed Modality Invariant (MI)-RAFT model architecture (top); Dual-pair synthetic training strategy with correlation lookup feature alignment (bottom left) and inference procedure of the proposed MI-RAFT model/fine alignment (bottom right).}
\label{fig:miraft}
\vspace{-0.1cm}
\end{figure*}

The architecture of the proposed MI-RAFT model is illustrated in
Fig.~\ref{fig:miraft}. MI-RAFT extends the correlation-feature branch of RAFT \cite{Correspondence-Teed-RAFT} by incorporating features from a pretrained modality-invariant vision model. Specifically, the source image $I_{\mathrm{src}}$ and target image $I_{\mathrm{tgt}}$ are processed by both the pretrained RAFT feature encoder and a pretrained modality-invariant feature extractor, producing flow-specific features $\mathbf{x}_{\mathrm{src}}^{\mathrm{flow}}$ and
$\mathbf{x}_{\mathrm{tgt}}^{\mathrm{flow}}$, and modality-invariant features $\mathbf{x}_{\mathrm{src}}^{\mathrm{MI}}$ and
$\mathbf{x}_{\mathrm{tgt}}^{\mathrm{MI}}$, respectively. We experiment with features extracted from both generative and discriminative pretrained models, namely DIFT \cite{Correspondence-DIFT-NeurIPS} and DINOv3 \cite{Others-DINOV3}, and observe similar downstream performance. Separate learnable adapters first refine and align the two feature types. Their outputs are then fused by element-wise addition and further processed by a post-fusion adapter. The resulting features are used to construct the all-pairs correlation volume employed by the recurrent RAFT flow estimator.

\begin{figure}[ht]
\centering
\includegraphics[width=0.49\textwidth]{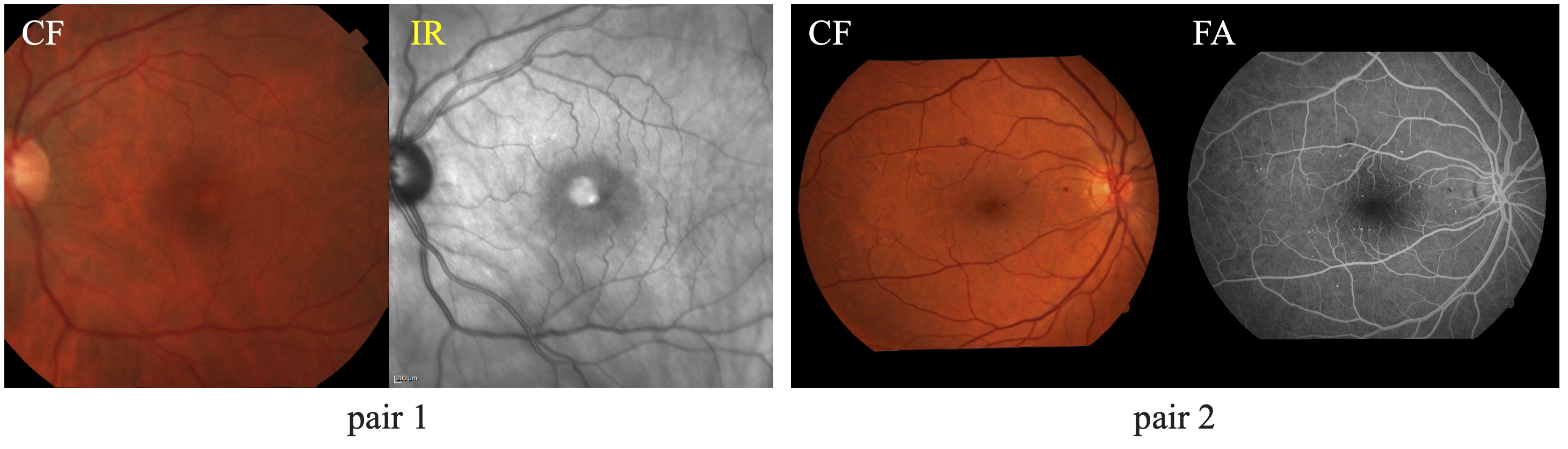}
\vspace{-0.8cm}
\caption{Two toy cross-modal retinal image pairs for Table~\ref{tab:miraft_explain_toy}.}
\label{fig:toy_pairs}
\end{figure}

To illustrate the complementary properties of the two feature types, we analyze two aligned cross-modal retinal image pairs, shown in
Fig.~\ref{fig:toy_pairs}, using a trained MI-RAFT model. Table \ref{tab:miraft_explain_toy} reports two attributes of the representations at different stages of the model. First, we measure cross-modal feature consistency by averaging the cosine similarity between source and target feature vectors at corresponding spatial locations. Because the image pairs are already aligned, a larger value indicates greater invariance to the modality-dependent appearance differences. Cross-modal consistency alone, however, does not guarantee reliable flow estimation. Because RAFT and related optical flow models infer displacement from the local structure of a cost or correlation volume \cite{Correspondence-Teed-RAFT,Correspondence-PWCNet-CVPR}, the underlying features must also distinguish a location from nearby, visually similar locations. We therefore use a self-correlation hard-negative margin as a proxy for local feature discriminability. For a feature map $\mathbf{f}$ with input-image stride $s$, we define
\begin{equation}
M_{\mathrm{HN}}(\mathbf{f}) =
\frac{1}{|\Omega|}
\sum_{x\in\Omega}
\left[
1-
\max_{\substack{
y\in\mathcal{N}_{R}^{(s)}(x)\\
y\notin\mathcal{E}_{\delta}^{(s)}(x)
}}
\cos\left(\mathbf{f}(x),\mathbf{f}(y)\right)
\right],
\end{equation}
where $\Omega$ contains all valid feature locations whose search regions lie entirely within the feature map. The search neighborhood and center-exclusion region are defined in input-image coordinates as
\begin{equation}
\mathcal{N}_{R}^{(s)}(x)=\left\{y:s\lVert y-x\rVert_{\infty}\le R\right\}
\end{equation}
\begin{equation}
\mathcal{E}_{\delta}^{(s)}(x)=\left\{y:s\lVert y-x\rVert_{\infty}\le\delta\right\}
\end{equation}
We use $R=64$ pixels and $\delta=16$ pixels for all representations, ensuring that features with different spatial strides are evaluated over the same physical regions. For each image pair, the reported value is the average of the margins computed independently on the source and target images. A larger $M_{\mathrm{HN}}$ indicates that each feature is more clearly separated from its most similar nearby nontrivial match and is therefore more likely to produce a sharp, unambiguous local correlation peak.

\begin{table}[ht]
    \vspace{-0.1cm}
    \caption{Toy examples on two real image pairs illustrating how feature characteristics evolve within MI-RAFT to improve both modality invariance and local discriminability of the correlation landscape. $M_{\mathrm{HN}}$ denotes the hard-negative margin, and cos-sim denotes cosine similarity. RAFT and DINO denote the raw outputs of the RAFT feature encoder and DINO feature extractor, respectively; Fused denotes the adapted fused features used to construct the correlation volume; and Fused-CLFA denotes the corresponding features from the model trained with the proposed Correlation-Lookup Feature Alignment (CLFA) strategy.}
    \label{tab:miraft_explain_toy}
    \centering

    \begin{tabular}{c|c|cccc}
        \toprule
         & Measures & RAFT & DINO & Fused & Fused-CLFA \\
        \midrule
        Toy Pair 1& $M_{HN}$ & 0.5808 & 0.0218 & 0.5288 & 0.5367 \\
        (Fig.\ref{fig:toy_pairs} L)& cos-sim & 0.1281 & 0.8707 & 0.2303 & 0.3003 \\
        \midrule
        Toy Pair 2& $M_{HN}$ & 0.5041 & 0.0232 & 0.4865 & 0.4743 \\
        (Fig.\ref{fig:toy_pairs} R)& cos-sim & 0.1189 & 0.8951 & 0.2707 & 0.3053 \\

        \bottomrule
    \end{tabular}
\end{table}

As shown in Table~\ref{tab:miraft_explain_toy}, the raw DINOv3 features provide strong cross-modal consistency but have very small hard-negative margins, indicating poor local discriminability. Conversely, the raw RAFT features are locally discriminative but exhibit low cross-modal similarity. The learned fused representation improves cross-modal similarity over the RAFT features while retaining most of their hard-negative margin.

The rest of the model architecture is the same as RAFT, which uses a context encoder with identical structure with the feature encoder to encode context information from $I_{tgt}$. Note that we are predicting the flow from $I_{tgt}$ to $I_{src}$, this is because for alignment we need the backward sampling grid which is constructed from the inverted flow. The constructed modality-invariant correlation volume, along with the context feature, are used by a GRU to iteratively predict and refine the flow. During training, we set both the context encoder and the GRU to be learnable. 

\subsection{Synthetic Pair Learning and Correlation Lookup Feature Alignment Strategy}
\label{subsec:training_strategy}
Because dense ground-truth flow is difficult to obtain for real cross-modal retinal image pairs, we train MI-RAFT using synthetically generated pairs and subsequently evaluate its generalization to real cross-modal pairs. Given a single retinal image, we first apply a small random affine transformation. We then sample a Gaussian noise field, smooth it using a Gaussian filter, and use the resulting displacement field as a spatially varying non-linear residual to the affine transformation. We refer to the combination of these global and local transformations as Geometric Synthesis ($GS$). Because both components are known, the ground-truth flow can be obtained directly from their composition. To avoid supervising invalid correspondences, we also generate a binary validity mask by warping the source field-of-view mask through the composed transformation and retaining only target pixels whose mapped source locations remain inside the valid retinal region.

Subsequently, independent random conventional intensity augmentations are first applied to the original and geometrically transformed images, which we term Intensity Synthesis ($IS$); implementation details are provided in Sec.~\ref{subsec:imple_details}. Additionally, pairs synthesized from the same retinal image can retain unrealistically strong appearance correspondence, we further apply independently sampled low-frequency illumination and focus/resolution perturbations to the two views. Given a normalized grayscale image $I\in[0,1]$, the illumination perturbation is defined as
\begin{equation}
\begin{aligned}
    \widetilde{I}_{\mathrm{ill}}(\mathbf{x})
    &=
    \operatorname{clip}\!\left(
    I(\mathbf{x})[1+\alpha L_g(\mathbf{x})]
    +\beta L_b(\mathbf{x}),\,0,\,1
    \right), \\
    L_k
    &=
    \operatorname{Norm_{(0,1)}}\!\left(
    \mathcal{U}^{\mathrm{bic}}_{H,W}(\epsilon_k)
    \right), \\
    &
    \epsilon_k\sim\mathcal{N}(0,1)^{
    \max(2,\lceil H/c_k\rceil)\times
    \max(2,\lceil W/c_k\rceil)},
\end{aligned}
\end{equation}
where $\operatorname{Norm_{(0,1)}}$ denotes zero-mean, unit-variance normalization; $\alpha$ and $\beta$ are randomly sampled gain and bias amplitudes; and $c_g$ and $c_b$ are the smooth-field scale parameter. The simulated focus/resolution perturbation is
\begin{equation}
    \widetilde{I}_{\mathrm{foc}}
    =
    G_{\sigma}*
    \mathcal{U}^{\mathrm{bic}}_{s}
    \left(\mathcal{D}^{\mathrm{area}}_{s}(I)\right),
\end{equation}
where the image is downsampled ($\mathcal{D}^{\mathrm{area}}_{s}(\cdot)$) by a random factor $s$, restored to its original resolution by bicubic interpolation ($\mathcal{U}^{\mathrm{bic}}_{s}(\cdot)$), and smoothed using a Gaussian kernel with randomly sampled standard deviation $\sigma$. Different from using Gaussian smoothing alone, the simulated focus/resolution change additionally applies downsampling and upsampling, explicitly reduces the effective spatial resolution and irreversibly suppresses fine-scale structures before Gaussian smoothing. Illumination perturbation modifies slowly varying background and contrast while preserving fine texture, whereas focus/resolution perturbation suppresses shared high-frequency details while retaining coarse intensity structure. We hypothesize that their joint application reduces these complementary shortcuts and encourages correspondence estimation from retinal structural geometry shared across modalities. Our experiments show that this strategy consistently improves registration performance.

The resulting images form a synthetic training pair whose analytically available flow directly supervises the iterative MI-RAFT predictions:
\begin{equation}
\mathcal{L}_{\mathrm{flow}}
=
\sum_{i=0}^{T-1}
\alpha^{T-i-1}
\frac{
\sum_{b,c,y,x}
m_{b,1,y,x}
\left(
\hat{u}_{i,b,c,y,x}-u_{b,c,y,x}
\right)^2
}{
2\sum_{b,y,x}m_{b,1,y,x}
},
\end{equation}
where $T$ is the total number of recurrent refinement iterations,
$\alpha=0.8$ controls the per-iteration loss weighting, $m$ is the binary validity mask, $\hat{u}_i$ is the flow predicted at iteration $i$, and $u$ is the ground-truth flow. The factor of $2$ in the denominator accounts for the horizontal and vertical flow components.

Furthermore, to discourage abrupt spatial variations that may produce unrealistic local deformations, we impose a first-order smoothing loss on the predicted
flow:
\begin{equation}
\mathcal{L}_{\mathrm{sm}}
=
\sum_{i=0}^{T-1}
\alpha^{T-i-1}
\left[
\ell_y(\hat{\mathbf{u}}_i,\mathbf{m})
+
\ell_x(\hat{\mathbf{u}}_i,\mathbf{m})
\right],
\end{equation}
where
\begin{equation}
\ell_y(\hat{\mathbf{u}},\mathbf{m})
=
\frac{
\sum_{b,c,y,x}
m^{y}_{b,y,x}
\left(
\hat{u}_{b,c,y+1,x}-\hat{u}_{b,c,y,x}
\right)^2
}{
2\sum_{b,y,x}m^{y}_{b,y,x}
},
\end{equation}
\begin{equation}
\ell_x(\hat{\mathbf{u}},\mathbf{m})
=
\frac{
\sum_{b,c,y,x}
m^{x}_{b,y,x}
\left(
\hat{u}_{b,c,y,x+1}-\hat{u}_{b,c,y,x}
\right)^2
}{
2\sum_{b,y,x}m^{x}_{b,y,x}
},
\end{equation}
with
\begin{equation}
\begin{aligned}
m^{y}_{b,y,x} &= m_{b,1,y,x}m_{b,1,y+1,x}, \\
m^{x}_{b,y,x} &= m_{b,1,y,x}m_{b,1,y,x+1}.
\end{aligned}
\end{equation}
Thus, each finite difference contributes to the loss only when both adjacent pixels have valid ground-truth correspondences.

Although feature fusion substantially improves cross-modal correspondence and already produces accurate flow estimates, the fused representation remains less modality-invariant than the raw DINOv3 representation, as illustrated in Table~\ref{tab:miraft_explain_toy}. We therefore introduce a dual-pair synthetic training strategy with Correlation-Lookup Feature Alignment (CLFA) to further improve modality invariance without directly constraining the fused descriptors used to construct the correlation volume.


As illustrated in the lower-left part of Fig.~\ref{fig:miraft}, we generate two synthetic training pairs from each individual retinal image. Both pairs share the same target image,
\begin{equation}
I_{\mathrm{tgt}} = IS_1(I),
\end{equation}
to which no geometric transformation is applied. Their source images are
generated using two independently sampled intensity transformations but the same geometric transformation:
\begin{equation}
I_{\mathrm{src}}^{A}=IS_2\!\left(GS_1(I)\right),
\qquad
I_{\mathrm{src}}^{B}=IS_3\!\left(GS_1(I)\right).
\end{equation}
Consequently, $I_{\mathrm{src}}^{A}$ and $I_{\mathrm{src}}^{B}$ have different synthetic modality appearances but identical spatial geometry. The two source--target pairs therefore share the same ground-truth flow:
\begin{equation}
\left(I_{\mathrm{src}}^{A},I_{\mathrm{tgt}}\right),
\qquad
\left(I_{\mathrm{src}}^{B},I_{\mathrm{tgt}}\right).
\end{equation}
Both pairs are processed by the same MI-RAFT model during each training
iteration.

In RAFT, the source and target features are first used to construct an
all-pairs correlation volume and its multi-scale correlation pyramid. At each recurrent refinement iteration, the current flow estimate specifies a candidate correspondence for every source-grid location. RAFT then samples correlation values from a local neighborhood around each candidate correspondence at multiple pyramid levels and concatenates them into a correlation-lookup feature, denoted by $\mathcal{C}_{p;i}$. This feature represents the local matching evidence retrieved from the correlation pyramid at iteration $i$ and is provided to the recurrent update block to predict the next flow refinement.

Because the two synthetic pairs differ only in their intensity transformations and share the same underlying geometry, their correlation-lookup features should provide consistent matching evidence at corresponding spatial locations. We therefore align these features using a mean $\ell_1$ loss:
\begin{equation}
\mathcal{L}_{\mathrm{CLFA}}
=
\sum_{i=0}^{T-1}
\frac{
\sum_{b,d,y,x}
m_{b,1,y,x}^{\mathcal{D}}
\left|
\mathcal{C}_{p;i,b,d,y,x}^{A}
-
\mathcal{C}_{p;i,b,d,y,x}^{B}
\right|
}{
D_{\mathcal{C}}
\sum_{b,y,x}m_{b,1,y,x}^{\mathcal{D}}
},
\end{equation}
where $d$ indexes the $D_{\mathcal{C}}$ channels of the correlation-lookup feature and $m^{\mathcal{D}}$ refers to the validity mask downsampled to the feature map size. Table~\ref{tab:miraft_explain_toy} shows that the feature modality invariance further improves after CLFA and our experiments show that this strategy allows consistent improvement in the final registration performance.

Finally, the proposed MI-RAFT model is jointly optimized by:
\begin{equation}
\mathcal{L}_{total}=\mathcal{L}_{flow} + \lambda_{sm}\mathcal{L}_{sm} + \lambda_{CLFA}\mathcal{L}_{CLFA}
\end{equation}
where $\lambda_{sm}$ and $\lambda_{CLFA}$ are the hyperparameters to control the weight for the smoothing and CLFA loss, respectively.

\subsection{Inference and Alignment}
At inference time, the inverse flow $\mathbf{u}(\mathbf{p})$ from $I_{tgt}$ to $I_{src}$ is predicted from the input images. Then the backward sampling grid $\mathbf{G}(\mathbf{p})$ used for image alignment is generated by adding an identity grid $\mathbf{I}(\mathbf{p})$ to the inverse flow:
\begin{equation}
\mathbf{G}(\mathbf{p})
=
\mathbf{I}(\mathbf{p}) + \mathbf{u}(\mathbf{p}),
\qquad
\mathbf{I}(\mathbf{p}) = (x,y)
\end{equation}
where $\mathbf{p}$ is each pixel location on the target image $I_{tgt}$.

\section{Experiments}
\label{sec:experiments}
\subsection{Datasets}
\subsubsection{Training}
To fine-tune the keypoint descriptor head of the coarse-alignment model on vessel maps generated by URVSM, we use the same training dataset as Wang et al. \cite{Registration-Wang-TIP}. The dataset comprises 530 paired Topcon color fundus (CF) and Heidelberg infrared reflectance (IR) images. Each image pair is annotated with at least ten point correspondences, which are used to estimate the ground-truth global transformation between the two images.

To train the fine-alignment MI-RAFT model, we construct an unpaired dataset of 2,500 retinal images, comprising 500 images from each of five modalities: CF, scanning laser ophthalmoscopy multicolor (MC), fundus autofluorescence (FAF), fluorescein angiography (FA), and infrared reflectance (IR). Synthetic training pairs with known ground-truth flow are generated from these individual images using the geometric and intensity synthesis procedures described in Sec.~\ref{subsec:training_strategy}. For selected ablation experiments, we use a smaller randomly selected subset of 500 images, containing 100 images from each modality.

\subsubsection{Evaluation}
We evaluate the proposed method on four diverse real-world retinal image registration datasets spanning both single- and cross-modality settings: (1) the FIRE dataset \cite{Dataset-FIRE}, which contains 134 pairs of color fundus (CF) images acquired using a Nidek camera system and are used to evaluate single-modality registration; (2) the CF--IR dataset \cite{Registration-Wang-TIP}, comprising 253 pairs of Topcon CF and Heidelberg infrared reflectance (IR) images; (3) the CF--FA dataset \cite{Dataset-CFFA}, comprising 29 pairs of Canon CF and fluorescein angiography (FA) images; and (4) our newly collected FAF--MC dataset, which contains 319 pairs of fundus autofluorescence (FAF) and multicolor SLO (MC) images acquired using an Optos camera system. The last dataset enables evaluation on a modality combination not covered by existing retinal registration datasets.

The original FAF and MC images have an ultra-wide field of view of approximately $135^{\circ}$. Since this work focuses on modality-invariant registration rather than registration across different fields of view, we crop both images in each pair to a central $30^{\circ}\times30^{\circ}$ region. This produces paired images with fields of view comparable to those of the other evaluation datasets and isolates the cross-modality registration problem considered in this work. All training and evaluation images are in resolution $768 \times768$.

\subsection{Implementation Details}
\label{subsec:imple_details}
For coarse alignment, we use the URVSM \cite{Segmentation_Wen_TIP} variant with CycleGAN \cite{Others-CycleGAN} as the image-translation backbone and ResDO-UNet \cite{Others_ResdoU-Net} as the segmentation backbone. We fine-tune the descriptor head for 300 epochs using an initial learning rate of $10^{-4}$, which is multiplied by 0.3 at epochs 50 and 125. The triplet loss \cite{Registration-SuperRetina} uses a margin of 0.8. To extract keypoints from the predicted probability map, we apply non-maximum suppression with a window size of 7 pixels and a detection threshold of 0.01. Initial sparse correspondences are established using mutual nearest-neighbor (MNN) matching \cite{Registration-Wang-TIP} with a similarity threshold of 0.9.

For fine alignment, we train MI-RAFT for 150 epochs using an initial learning rate of $10^{-4}$, which is multiplied by 0.3 at epoch 100. For the DINOv3-based variant, we use the ViT-B model from DINOv3 \cite{Others-DINOV3} and extract the final layer-normalized patch tokens. For the DIFT-based variant, following \cite{Correspondence-DIFT-NeurIPS,Correspondence_MIFNet_TIP}, we use Stable Diffusion v2.1 \cite{Others-stable_diffusion} and extract features at denoising step 60 from the second upsampling block above the bottleneck of the denoising U-Net. Given the computational cost of processing dense image features, we use a lightweight double-convolution block (Fig.~\ref{fig:miraft}) as an efficient feature adapter. In the modality-invariant feature adapter, a $2\times$ upsampling operation is inserted between the two blocks to increase the spatial resolution of the stride-16 DINOv3 or DIFT features to the stride-8 resolution used by RAFT. The remaining RAFT components follow the default configuration of the original model \cite{Correspondence-Teed-RAFT}.

\begin{table*}[ht]
  \caption{Quantitative comparison of coarse (sparse-feature-based) alignment with state-of-the-art methods. ORN denotes the same pretrained outlier-rejection network used in our method \cite{Registration-Wang-TIP,Correspondence_CLe_CVPR}, and SG denotes SuperGlue \cite{Correspondence-SuperGlue-CVPR}. Unless otherwise specified, each baseline uses the default feature matcher/outlier rejection from its original implementation. Results are reported as mean $\pm$ standard deviation.}
  \vspace{-0.1cm}
  \label{tab:main_tab_coarse}
  \centering
  \scriptsize

  \begin{tabular}{c|cc|cc|cc|cc}
  \toprule
  \multirow{2}{*}{Method/Dataset} & \multicolumn{2}{c|}{FIRE (CF-CF)} & \multicolumn{2}{c|}{CF-IR} & \multicolumn{2}{c|}{CF-FA} & \multicolumn{2}{c}{FAF-MC} \\
  \cmidrule{2-9}
   & $Dice$ & $Dice_{s}$ & $Dice$ & $Dice_{s}$ & $Dice$ & $Dice_{s}$ & $Dice$ & $Dice_{s}$ \\
  \midrule
   SuperPoint \cite{Registration-SuperPoint} & 0.2593($\pm$0.1663) & 0.2915 & 0.0397($\pm$0.0304) & 0.0602 & 0.1121($\pm$0.0373) & 0.2139 & 0.0967($\pm$0.0458) & 0.1314 \\
   SuperPoint+SG \cite{Correspondence-SuperGlue-CVPR} & 0.6190($\pm$0.1926) & 0.6317 & 0.1653($\pm$0.1801) & 0.2026 & 0.2997($\pm$0.1665) & 0.3960 & 0.4347($\pm$0.1793) & 0.4690 \\
   SuperPoint+ORN \cite{Registration-SuperPoint, Correspondence_CLe_CVPR} & 0.6542($\pm$0.1906) & 0.6611 & 0.4442($\pm$0.2644) & 0.4626 & 0.4071($\pm$0.2502) & 0.4916 & 0.5239($\pm$0.1401) & 0.5499 \\
   Y. Wang et al. \cite{Registration-Wang-TIP} & 0.6200($\pm$0.1564) & 0.6332 & 0.5724($\pm$0.1837) & 0.6091 & 0.4257($\pm$0.1665) & 0.5091 & 0.3715($\pm$0.1963) & 0.4057 \\
   SuperRetina \cite{Registration-SuperRetina} & 0.4745($\pm$0.2534) & 0.4957 & 0.3691($\pm$0.2474) & 0.3985 & 0.0871($\pm$0.0542) & 0.1647 & 0.3417($\pm$0.1892) & 0.3779 \\
   SuperRetina+ORN \cite{Registration-SuperRetina, Correspondence_CLe_CVPR} & 0.6687($\pm$0.1984) & 0.6710 & 0.3832($\pm$0.2517) & 0.4112 & 0.3018($\pm$0.1408) & 0.3913 & 0.5174($\pm$0.1372) & 0.5455 \\
   SuperJunction \cite{Registration-SuperJunction} & 0.6567($\pm$0.2133) & 0.6626 & 0.0887($\pm$0.0286) & 0.1278 & 0.0622($\pm$0.0299) & 0.1419 & 0.0663($\pm$0.0212) & 0.0993 \\
   RetinaRegNet-DIFT \cite{Registration-Sivaraman-RetRegNet} & 0.4286($\pm$0.2583) & 0.4547 & 0.0712($\pm$0.0285) & 0.1080 & 0.1609($\pm$0.1089) & 0.2666 & 0.1879($\pm$0.1309) & 0.2255 \\
   MIF-Net \cite{Correspondence_MIFNet_TIP} & 0.6466($\pm$0.1922) & 0.6546 & 0.4004($\pm$0.1717) & 0.4340 & 0.5337($\pm$0.0893) & 0.6123 & 0.4842($\pm$0.1152) & 0.5175 \\
   MINIMA-LG \cite{Correspondence-Ren-MINIMA} & 0.5647($\pm$0.2320) & 0.5808 & 0.5579($\pm$0.2022) & 0.5748 & 0.5952($\pm$0.1311) & 0.6629 & \textbf{0.5412($\pm$0.1216)} & 0.5622 \\
   EyeKey \cite{Registration-Liang-EyeKey} & 0.5971($\pm$0.2764) & 0.6045 & 0.0727($\pm$0.0371) & 0.1162 & 0.0806($\pm$0.0414) & 0.1940 & 0.0661($\pm$0.0380) & 0.1132 \\

   \textbf{URVSM-Reg (ours)} & \textbf{0.6688($\pm$0.1943)} & \textbf{0.6866} & \textbf{0.6005($\pm$0.1667)} & \textbf{0.6324} & \textbf{0.6228($\pm$0.0928)} & \textbf{0.7103} & 0.5379($\pm$0.1249) & \textbf{0.5634} \\
  \bottomrule
  \end{tabular}
\end{table*}

\begin{figure*}[ht]
\centering
\includegraphics[width=0.98\textwidth]{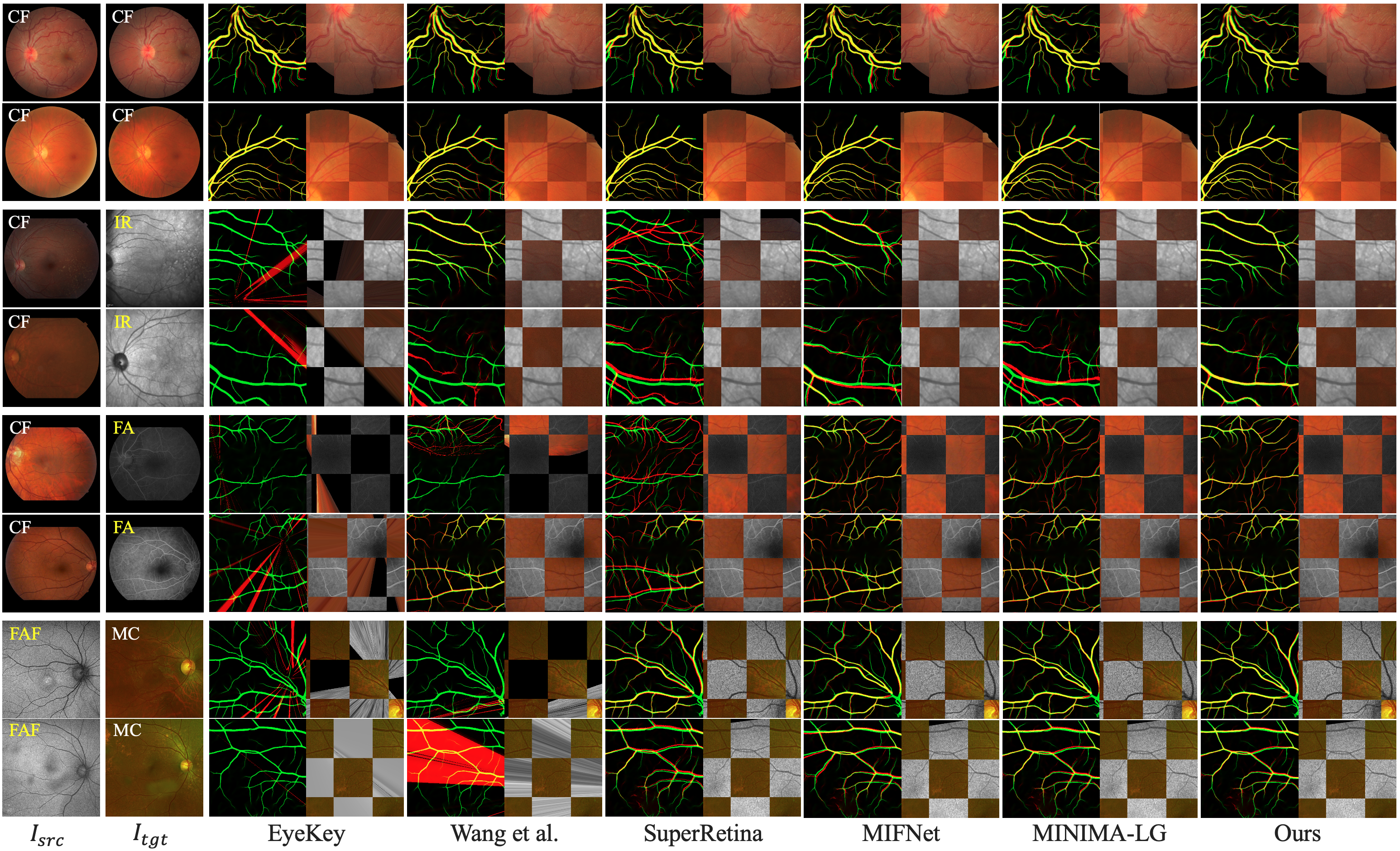}
\vspace{-0.1cm}
\caption{Qualitative comparison with state-of-the-art methods in coarse global modality-invariant retinal image registration. On the vessel overlap maps, yellow shows overlay-ed vessels, red and green shows non-overlayed vessels from source and target images, respectively.}
\label{fig:qualitative_coarse}
\vspace{-0.1cm}
\end{figure*}

To generate synthetic training pairs, we independently apply random brightness and contrast adjustments, Gaussian noise, Gaussian blur, intensity inversion (negative-film transformation). For the low-frequency illumination perturbation, the smooth-field scale parameters $c_g$ and $c_b$ are sampled from [80, 140] and [96, 160] pixels while the gain weight $\alpha$ and bias weight $\beta$ are sampled from [0.08, 0.24] and [0.02, 0.08], respectively. Each augmentation is independently applied to an image with a probability of 0.3. For geometric synthesis, the random affine transformation is constrained to a rotation of at most $\pm3^{\circ}$, a translation of at most 10 pixels along each axis, a scaling variation of at most $\pm2\%$, and no shearing. The non-linear residual displacement is generated by sampling a two-channel Gaussian noise field $\boldsymbol{\epsilon}\in\mathbb{R}^{H\times W\times2}$, smoothing each channel with a Gaussian filter whose standard deviation is sampled as $\sigma\sim\mathcal{U}(20,30)$, and scaling the resulting field by $\lambda=10^r$, where $r\sim\mathcal{U}(-2,0)$. Here, $\mathcal{U}$ denotes the continuous uniform distribution. The loss weights are set to $\lambda_{\mathrm{sm}}=0.5$ and $\lambda_{\mathrm{CLFA}}=2\times10^{-4}$.

The proposed framework is implemented in Python using PyTorch. Final image warping is performed using the \texttt{torch.nn.functional.grid\_sample} function, which is an equivalence of the Spatial Transformer Network \cite{Registration-STN}. All experiments are conducted on a workstation equipped with an Intel i9-13900K CPU and NVIDIA RTX 4090 GPUs.

\begin{table*}[ht]
  \caption{Quantitative comparison of fine (dense-feature/optical-flow-based) alignment with state-of-the-art methods.}
  \vspace{-0.1cm}
  \label{tab:main_tab_fine}
  \centering
  \scriptsize

  \begin{tabular}{c|cc|cc|cc|cc}
  \toprule
  \multirow{2}{*}{Method/Dataset} & \multicolumn{2}{c|}{FIRE (CF-CF)} & \multicolumn{2}{c|}{CF-IR} & \multicolumn{2}{c|}{CF-FA} & \multicolumn{2}{c}{FAF-MC} \\
  \cmidrule{2-9}
  & $Dice$ & $Dice_{s}$ & $Dice$ & $Dice_{s}$ & $Dice$ & $Dice_{s}$ & $Dice$ & $Dice_{s}$ \\
  \midrule 
  Coarse Registration & 0.6688($\pm$0.1943) & 0.6866 & 0.6005($\pm$0.1667) & 0.6324 & 0.6228($\pm$0.0928) & 0.7103 & 0.5379($\pm$0.1249) & 0.5634 \\
  
  \midrule
  MI-B-Splines \cite{Registration-Klein-TIP, Registration-Klein-Elastix} & 0.7022($\pm$0.1047) & 0.7110 & 0.6153($\pm$0.1712) & 0.6345 & 0.6529($\pm$0.1087) & 0.7294 & 0.5578($\pm$0.1300) & 0.5865 \\
  RAFT \cite{Correspondence-Teed-RAFT} & 0.7559($\pm$0.1189) & 0.7657 & 0.6162($\pm$0.1554) & 0.6406 & 0.6373($\pm$0.1012) & 0.7166 & 0.5700($\pm$0.1100) & 0.6093 \\
  CrossRAFT \cite{Registration-CrossRAFT} & 0.7246($\pm$0.1481) & 0.7342 & 0.6225($\pm$0.1692) & 0.6421 & 0.6395($\pm$0.1076) & 0.7242 & 0.5697($\pm$0.1118) & 0.6079 \\
  Seg-DeformNet \cite{Registration-Zhang-TIP} & 0.6915($\pm$0.1588) & 0.7067 & 0.6531($\pm$0.1726) & 0.6638 & 0.4162($\pm$0.0907) & 0.5220 & 0.5503($\pm$0.1158) & 0.5824 \\
  MCM-RAFT \cite{Registration-MCMRAFT} & 0.7078($\pm$0.1497) & 0.7213 & 0.6111($\pm$0.1671) & 0.6328 & 0.6088($\pm$0.0991) & 0.6963 & 0.5663($\pm$0.1145) & 0.5968 \\
  GAMorph \cite{Registration-Liu-GAMorph} & 0.7405($\pm$0.1171) & 0.7455 & 0.6431($\pm$0.1537) & 0.6501 & 0.1211($\pm$0.0308) & 0.2639 & 0.6135($\pm$0.1058) & 0.6371 \\
  MINIMA-RoMA \cite{Correspondence-Ren-MINIMA} & 0.7751($\pm$0.1139) & 0.7759 & 0.6626($\pm$0.1473) & 0.6779 & 0.6613($\pm$0.0883) & 0.7550 & 0.6416($\pm$0.0994) & 0.6598 \\
  CRFT \cite{Registration-Liu-CRFT} & 0.6214($\pm$0.1601) & 0.6717 & 0.5761($\pm$0.1597) & 0.6172 & 0.5105($\pm$0.1055) & 0.6524 & 0.4842($\pm$0.1077) & 0.5436 \\
  
  \textbf{MI-RAFT-DIFT (ours)} & 0.7811($\pm$0.1136) & 0.7810 & 0.6694($\pm$0.1462) & 0.6821 & 0.6827($\pm$0.0883) & 0.7648 & 0.6562($\pm$0.0973) & 0.6698 \\
  \textbf{MI-RAFT-DINO (ours)} & \textbf{0.7827($\pm$0.1133)} & \textbf{0.7821} & \textbf{0.6701($\pm$0.1457)} & \textbf{0.6825} & \textbf{0.6835($\pm$0.0881)} & \textbf{0.7658} & \textbf{0.6569($\pm$0.0997)} & \textbf{0.6733} \\
  \bottomrule
  \end{tabular}
\end{table*}

\begin{figure*}[ht]
\centering
\includegraphics[width=0.98\textwidth]{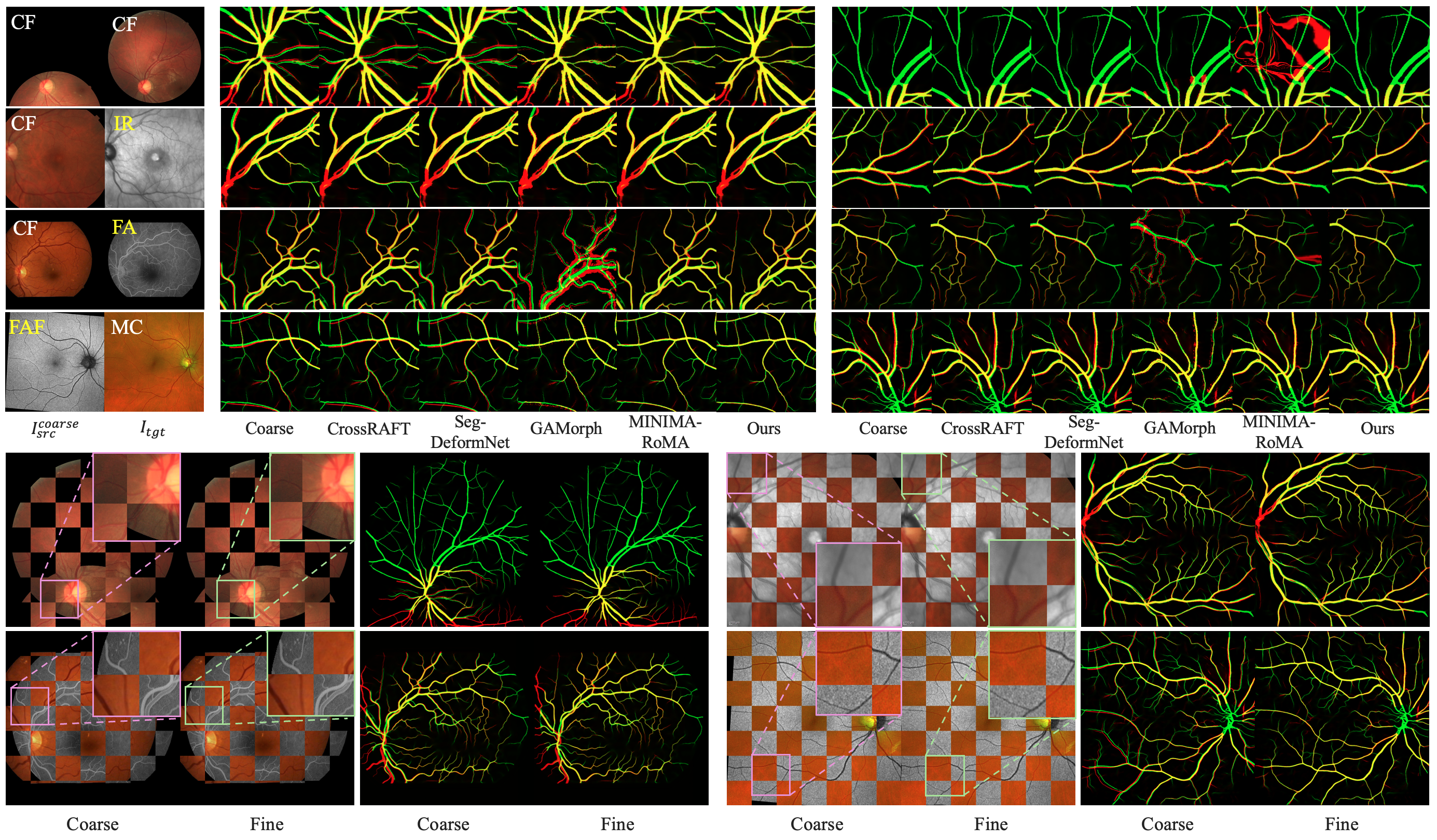}
\vspace{-0.2cm}
\caption{Qualitative comparison of fine (dense-correspondence- and optical-flow-based) alignment with state-of-the-art methods and comparison with the proposed coarse alignment. All methods are initialized from the same coarse alignment produced by our proposed coarse-registration method, where $I_{\mathrm{src}}^{\mathrm{coarse}}$ denotes the resulting coarsely aligned source image.}
\label{fig:qualitative_fine}
\vspace{-0.1cm}
\end{figure*}

\subsection{Evaluation Metrics}
We use the vessel-overlap-based Dice coefficient and soft Dice coefficient ($\mathrm{Dice}_{s}$) as common evaluation metrics for both coarse global alignment and fine local alignment. Prior research \cite{Registration-Wang-Metrics} shows that vessel-based binary and soft Dice coefficients exhibited the strongest correlations with subjective registration-quality assessments by retinal imaging experts among the evaluated objective metrics. These metrics have also been widely adopted in previous retinal image registration studies \cite{Registration-Wang-TIP,Registration-Zhang-TIP,Registration-Wen-ICIP,Registration-Kalaw-Eye}. Moreover, they do not require manually annotated point correspondences, enabling consistent evaluation across all datasets considered in this work. Specifically, we use the vessel maps generated by URVSM \cite{Segmentation_Wen_TIP} for computing both $\mathrm{Dice}$ and $\mathrm{Dice}_{s}$ in all experiments.

\subsection{Comparison with the State of the Art: Coarse Registration}
Quantitative and qualitative comparisons are presented in Table~\ref{tab:main_tab_coarse} and Fig.~\ref{fig:qualitative_coarse}, respectively. For coarse alignment, we compare our method with generic sparse correspondence methods \cite{Registration-SuperPoint,Correspondence-SuperGlue-CVPR,Correspondence_CLe_CVPR}, retinal-specific global registration methods \cite{Registration-Wang-TIP,Registration-SuperRetina,Registration-SuperJunction,Registration-Sivaraman-RetRegNet,Registration-Liang-EyeKey}, and recent modality-invariant correspondence methods fine-tuned on retinal imagery \cite{Correspondence_MIFNet_TIP,Correspondence-Ren-MINIMA}. Most competing methods perform well on either the single-modality CF dataset or the CF--IR setting represented in their training data \cite{Registration-Wang-TIP}, but generalize poorly to other modality combinations. The two modality-invariant methods exhibit broader generalization, although their performance remains inconsistent across datasets. In contrast, our method uses a single model without dataset-specific fine-tuning and achieves the strongest overall performance across all evaluated datasets and modality combinations.

\begin{figure*}[ht]
\centering
\includegraphics[width=1\textwidth]{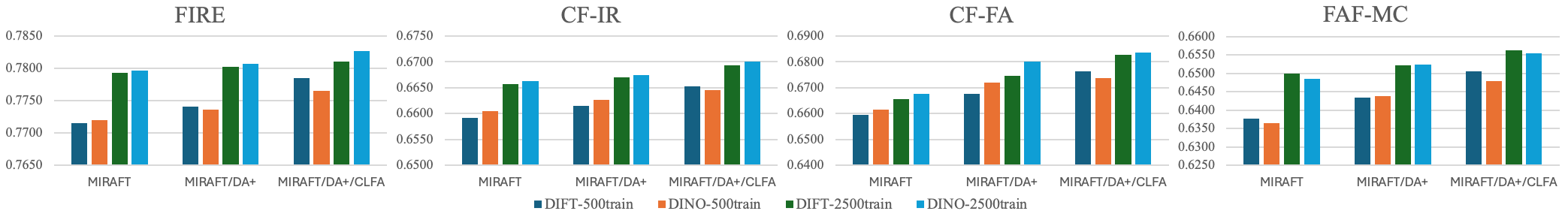}
\vspace{-0.7cm}
\caption{Ablation study of additional training strategies for MI-RAFT. DA+ denotes the  low-frequency illumination and focus/resolution perturbations, while CLFA denotes the proposed Correlation-Lookup Feature Alignment. Results are reported as mean Dice scores.}
\label{fig:ablation_learning}
\vspace{-0.3cm}
\end{figure*}

\begin{figure*}[ht]
\centering
\includegraphics[width=1\textwidth]{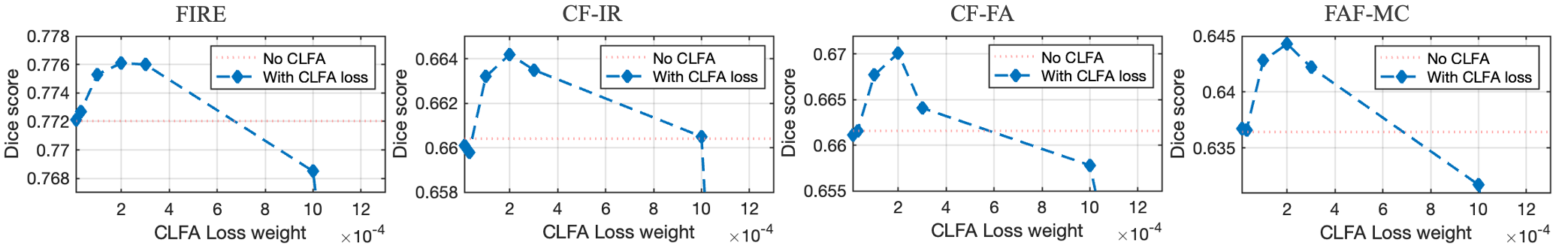}
\vspace{-0.7cm}
\caption{Ablation study of the effect of the CLFA loss weight $\lambda_{\mathrm{CLFA}}$ on final registration performance.}
\label{fig:ablation_weights}
\vspace{-0.3cm}
\end{figure*}

\subsection{Comparison with the State of the Art: Fine Registration}
All methods in this comparison take images aligned by our coarse-registration model as input. Quantitative and qualitative results are presented in Table~\ref{tab:main_tab_fine} and Fig.~\ref{fig:qualitative_fine}, respectively. We compare our method with a classical B-spline-based deformable registration method \cite{Registration-Klein-TIP,Registration-Klein-Elastix}, the generic RAFT model \cite{Correspondence-Teed-RAFT}, state-of-the-art cross-modal optical-flow methods \cite{Registration-CrossRAFT,Registration-MCMRAFT,Registration-Liu-CRFT}, retina-specific fine-registration methods \cite{Registration-Zhang-TIP,Registration-Liu-GAMorph}, and the modality-invariant dense correspondence method MINIMA-RoMA \cite{Correspondence-Ren-MINIMA}. Because RAFT and the cross-modal optical-flow models were not originally trained on retinal images, we fine-tune them using the same training images and synthetic-pair generation procedure as our MI-RAFT model, and use the reported best training hyperparameters in original papers to enable a fair and controlled comparison.

Although the cross-modal optical-flow methods generally improve upon the initial coarse alignment, their performance is considerably less reliable across datasets. In particular, the relatively coarse spatial feature resolution of CRFT \cite{Registration-Liu-CRFT} is unsuitable for retinal registration involving many fine image structures, resulting in performance that can be worse than the initial coarse alignment. Under the controlled fine-tuning setting, these comparisons demonstrate the advantage of the proposed MI-RAFT architecture. The retina-specific methods are more modality-dependent and generalize poorly to unseen modalities and modality combinations. MINIMA-RoMA produces more consistent results across modality combinations but can introduce substantial local distortions, particularly in regions where the source and target fields of view do not overlap. These artifacts may have only a limited effect on the Dice and soft-Dice scores, reflecting a known limitation of landmark overlap-based registration metrics \cite{Registration-Song-Thesis}; nevertheless, they remain undesirable for image registration. In contrast, our method achieves consistently more accurate fine alignment while exhibiting no visually apparent local distortion.

\subsection{Ablation Study}
\subsubsection{Additional Training Strategies for MI-RAFT}
As shown in Fig.~\ref{fig:ablation_learning}, we evaluate two additions to MI-RAFT training: (1) the low-frequency illumination and focus/resolution perturbations introduced into Intensity Synthesis ($IS$), beyond conventional intensity augmentations, and (2) the proposed dual-pair Correlation-Lookup Feature Alignment (CLFA) strategy. Although the baseline MI-RAFT already achieves highly accurate alignment, both strategies yield consistent improvements across different training-set sizes, evaluation datasets, and modality-invariant feature types (DIFT and DINOv3). The absolute Dice-score gains are relatively modest because most baseline alignments are already nearly perfect; consequently, the remaining disagreement primarily reflects differences between the vessel maps extracted from $I_{\mathrm{src}}$ and $I_{\mathrm{tgt}}$, rather than residual geometric misalignment.

Interestingly, the intensity-synthesis strategies and proposed CLFA also improve performance on the single-modality CF dataset. Although both images are CF, they can still exhibit considerable appearance variation due to differences in illumination, focus, resolution, color response, and acquisition conditions. By encouraging matching evidence to remain consistent under such variations, these strategies reduce reliance on appearance-specific cues and improve geometry-driven correspondence estimation. Thus, the learned appearance invariance benefits both cross-modality registration and intra-modality registration under acquisition-induced domain shifts.

\subsubsection{Effect of the CLFA Loss Weight}
All ablations in the following subsections use the 500-image training subset and exclude the low-frequency illumination and focus/resolution perturbations. Moreover, they uniformly use DINO as the modality-invariant feature. Fig.~\ref{fig:ablation_weights} examines the effect of the CLFA loss weight $\lambda_{\mathrm{CLFA}}$ on registration performance. CLFA consistently improves upon the baseline MI-RAFT model when $\lambda_{\mathrm{CLFA}}$ is between $10^{-4}$ and $3\times10^{-4}$, with the best performance achieved at $2\times10^{-4}$. However, further increasing the weight degrades alignment accuracy. These results indicate that CLFA is most effective as a softly weighted auxiliary regularizer; excessive weighting can overconstrain the correlation-lookup features and interfere with the primary flow-estimation objective.

\begin{table}[ht]
    \vspace{-0.1cm}
    \caption{Comparison of alternative feature-alignment strategies against the proposed CLFA method using DINOv3 as the modality-invariant feature extractor.}
    \label{tab:ablation_alter_feature_alignment}
    \centering

    \begin{tabular}{c|cccc}
        \toprule
         Dice & FIRE & CF-IR & CF-FA & FAF-MC \\
        \midrule
        Vanilla MI-RAFT & 0.7715 & 0.6604 & 0.6616 & 0.6364  \\
        GMM Clusering \cite{Correspondence_MIFNet_TIP} & 0.7592 & 0.6531 & 0.6546  & 0.6216  \\
        Pre-Corr Alignment & 0.7632 & 0.6559 & 0.6486 & 0.6203  \\
        \textbf{CLFA} & \textbf{0.7761} & \textbf{0.6642} & \textbf{0.6701} & \textbf{0.6443} \\
        \bottomrule
    \end{tabular}
    \vspace{-0.1cm}
\end{table}

\subsubsection{Alternative Feature-Alignment/Refinement Strategies}
In this section, we explore other feature-alignment/refinement strategies compared with our proposed CLFA. For each method, results from best finetuned loss weight is reported. 

As shown in Table~\ref{tab:ablation_alter_feature_alignment}, we first apply the Gaussian mixture model (GMM) clustering strategy introduced in \cite{Correspondence_MIFNet_TIP} to the dense, adapted modality-invariant features before fusion. However, this strategy does not transfer effectively to dense flow features: it performs worse than vanilla RAFT while increasing the training time by approximately $10\times$.

We also evaluate an auxiliary cosine feature-alignment loss applied to the adapted post-fusion features before constructing the correlation volume. Given a pair ($I_{src}, I_{tgt}$), for each valid target location, its corresponding source feature is bilinearly sampled using the ground-truth flow. The two features are then $\ell_2$-normalized, and their cosine distance is minimized over all valid correspondences. Although this objective encourages matched cross-modal locations to have similar fused representations, directly constraining the pre-correlation features substantially alters the statistics and matching structure of the resulting correlation scores, degrading the final registration performance. In contrast, CLFA aligns the task-relevant local matching evidence retrieved from the correlation pyramid without directly constraining the underlying features. These results demonstrate the effectiveness of the proposed CLFA design.

\begin{table}[ht]
    \vspace{-0.1cm}
    \caption{Ablation study of where to fuse the RAFT and modality-invariant (MI) features. ACBP denotes fusion after computing the top-level correlation maps but before downsampling them to construct the full correlation pyramid.}
    \label{tab:ablation_fusion_location}
    \centering

    \begin{tabular}{c|cccc}
        \toprule
        Dice & FIRE & CF-IR & CF-FA & FAF-MC \\
        \midrule
        Proposed Design & 0.7715 & \textbf{0.6604} & \textbf{0.6616} & \textbf{0.6364}  \\
        ACBP & 0.7696 & 0.6583 & 0.6550  & 0.6284  \\
        Corr. Lookup & \textbf{0.7730} & 0.6601 & 0.6608 & 0.6353  \\
        \bottomrule
    \end{tabular}
    \vspace{-0.1cm}
\end{table}

\subsubsection{Where to Fuse the RAFT and MI Features?}
We further investigate where the flow-specific RAFT features and modality-invariant (MI) features should be fused. All variants in this ablation are trained without CLFA to isolate the effect of the fusion location. As shown in Table~\ref{tab:ablation_fusion_location}, fusing the two top-level correlation volumes before constructing the multi-scale correlation pyramid, even with learned pre- and post-fusion adapters, notably degrades registration performance. Alternatively, fusing the correlation-lookup features retrieved from separate RAFT and MI correlation pyramids within the recurrent update block achieves performance comparable to our proposed design. However, this alternative requires constructing and querying two separate correlation pyramids, substantially increasing the computational cost. These results support our proposed fusion of RAFT and MI features before correlation computation as the most effective and computationally efficient design.

\begin{table}[ht]
    \vspace{-0.1cm}
    \caption{Ablation study on different choices of outlier rejection method in the proposed modality-invariant coarse registration model.}
    \label{tab:ablation_OR_method}
    \centering

    \begin{tabular}{c|cccc}
        \toprule
         Dice & FIRE & CF-IR & CF-FA & FAF-MC \\
        \midrule
        Pretrained CLe \cite{Registration-Wang-TIP, Correspondence_CLe_CVPR} & 0.6688 & \textbf{0.6005} & \textbf{0.6228} & \textbf{0.5379}  \\
        Finetuned CLe \cite{Registration-Wang-TIP, Correspondence_CLe_CVPR} & \textbf{0.6689} & 0.5990 & 0.6210 & 0.5367  \\
        Finetuned SuperGlue \cite{Correspondence-SuperGlue-CVPR} & 0.6510 & 0.5813 & 0.6037 & 0.5155  \\
        \bottomrule
    \end{tabular}
    \vspace{-0.1cm}
\end{table}

\subsubsection{Choice of Outlier-Rejection Model for Coarse Alignment}
To justify the use of the pretrained outlier-rejection network in Sec.~\ref{sec:coarse}, we evaluate several alternative training and matching strategies. First, following \cite{Registration-Wang-TIP}, we further fine-tune the outlier-rejection network using correspondences produced by the current coarse-alignment model. As shown in Table~\ref{tab:ablation_OR_method}, this additional fine-tuning does not improve registration performance. We also evaluate SuperGlue \cite{Correspondence-SuperGlue-CVPR} as a representative attention-based matcher. Because fully supervised SuperGlue training requires ground-truth assignments for the complete set of detected features, we adopt the weakly supervised adaptation strategy of \cite{Application-eye_tracking}. Nevertheless, SuperGlue does not outperform the more lightweight CLe architecture \cite{Correspondence_CLe_CVPR} employed in our framework.

\subsection{Runtime Analysis}
We report the inference time (per image pair) of the proposed coarse- and fine-registration methods in Table~\ref{tab:runtime}. Runtime is measured on the FAF-MC dataset using a single GPU on the hardware specified in Sec.~\ref{subsec:imple_details}. All input images have a resolution of $768\times768$ pixels.

\begin{table}[ht]
    \vspace{-0.1cm}
    \caption{Inference time of the proposed modality-invariant coarse and fine retinal image registration algorithms.}
    \label{tab:runtime}
    \centering

    \begin{tabular}{cc|c}
        \toprule
         \multicolumn{2}{c|}{} & Inference Time (s)\\
        \midrule
        \multirow{2}{*}{Coarse} & Segmentation (URVSM) & 0.0238($\pm$0.0162)   \\
         & Registration & 0.0308($\pm$0.0195)  \\
         \midrule
         \multicolumn{2}{c|}{Fine Registration (MI-RAFT-DINO)} & 0.1095($\pm$0.0147)  \\
         \multicolumn{2}{c|}{Fine Registration (MI-RAFT-DIFT)} & 0.2034($\pm$0.0180)  \\
        \bottomrule
    \end{tabular}
    \vspace{-0.1cm}
\end{table}

\section{Conclusions}
\label{sec:conclusions}

\subsection{Conclusion}
\label{ssec:conclusion}
In this work, we presented a modality-invariant coarse-to-fine framework for retinal image registration. For coarse global alignment, a universal retinal vessel segmentation model serves as a modality-invariant domain adapter, allowing a single sparse correspondence model to generalize across retinal modalities and modality combinations. For fine local alignment, we introduced MI-RAFT, which integrates flow-optimized and modality-invariant features for iterative dense correspondence estimation. Its synthetic-pair generation and CLFA strategies enable supervised flow learning from unpaired retinal images while further improving robustness to modality- and acquisition-dependent appearance variations.

Experiments on four diverse single- and cross-modality datasets, including our newly collected FAF--MC dataset, demonstrate that the proposed framework consistently outperforms state-of-the-art retinal registration, modality-invariant correspondence, and cross-modal optical-flow methods without dataset- or modality-pair-specific fine-tuning. By reducing the need for paired training data and separate models for different modality combinations, this work provides a more generalizable foundation for multimodal information co-localization, longitudinal comparison, and other retinal image analysis applications.

\subsection{Limitations and Future Work}
\label{ssec:limitations and future work}
The current framework assumes comparable fields of view and sufficient anatomical overlap, and therefore does not directly address extreme field-of-view differences. Its performance may also depend on vessel-segmentation quality and the realism of synthetic deformations. Future work will jointly address modality and field-of-view variations in the registration, moving toward fully universal retinal image registration.

\bibliographystyle{IEEEtran}
\bibliography{ref}

@article{Intro-medical_background,
  title={American Society of Retina Specialists Clinical Practice Guidelines on Multimodal Imaging for Retinal Disease},
  author={M. Ramakrishnan and J. Kovach and C. Wykoff and A Berrocal and Y. Modi},
  journal={Journal of VitreoRetinal Diseases},
  volume={8},
  number={3},
  pages={234-246},
  year={2024},
}

@article{Intro-coarse2fine_background,
  title={Medical image registration and its application in retinal images: a review},
  author={Q. Nie and X. Zhang and Y. Hu and M. Gong and J. Liu},
  journal={Visual Computing for Industry, Biomedicine, and Art},
  volume={7},
  number={1},
  pages={21},
  year={2024},
}

@article{Registration-Ding-coarse2fine,
  author={L. Ding and T. Kang and A. Kuriyanand R. Ramchandran and C. Wykoffand and G. Sharma},
  journal={IEEE Transactions on Biomedical Engineering}, 
  title={Combining Feature Correspondence With Parametric Chamfer Alignment: Hybrid Two-Stage Registration for Ultra-Widefield Retinal Images}, 
  year={2023},
  volume={70},
  number={2},
  pages={523-532},
}

@article{Registration-Zhang-TIP,
  title={Two-Step Registration on Multi-Modal Retinal Images via Deep Neural Networks},
  author={J. Zhang and Y. Wang and J. Dai and M. Cavichini and D.-U.G.Bartsch and W. R. Freeman and T. Q. Nguyen and C. An},
  journal={IEEE Transactions on Image Processing},
  volume={31},
  pages={823-838},
  year={2022},
}

@article{Registration-Wang-TIP,
  title={Robust Content-Adaptive Global Registration for Multimodal Retinal Images Using Weakly Supervised Deep-Learning Framework},
  author={Y. Wang and J.Zhang and M. Cavichini and D. Bartsch and W. Freeman and T. Nguyen and C. An},
  journal={IEEE Transactions on Image Processing},
  volume={30},
  pages={3167-3178},
  year={2021},
}

@inproceedings{Registration-Wang-ICASSP,
   author={Y. Wang and J. Zhang and C. An and M. Cavichini and M. Jhingan and M. Amador-Patarroyo and C. Long and D. Bartsch and W. Freeman and T. Nguyen},
   year={2020},
   title={A segmentation based robust deep learning framework for multi-modal retinal image registration},
   booktitle={2020 IEEE International Conference on Acoustics, Speech and Signal Processing (ICASSP)},
   pages={1369-1373},
}

@inproceedings{Registration-SuperRetina,
   author={Jiazhen Liu and Xirong Li and Qijie Wei and Jie Xu and Dayong Ding },
   year={2022},
   title={Semi-supervised Keypoint Detector and Descriptor for Retinal Image Matching},
   booktitle={2022 European Conference on Computer Vision (ECCV)},
   pages={593-609},
}

@inproceedings{Registration-Mahapatra,
   author={D. Mahapatra and B. Antony and S. Sedai and R. Garnavi},
   year={2018},
   title={Deformable medical image registration using generative adversarial networks},
   booktitle={IEEE 15th Int. Symp. Biomed. Imag. (ISBI)},
   pages={1449–1453},
}

@article{Registration-Luo,
  title={Multimodal affine registration for ICGA and MCSL fundus images of high myopia},
  author={G. Luo et al.},
  journal={Biomed. Opt. Exp.},
  volume={11},
  number={8},
  pages={4443-4457},
  year={2020},
}

@inproceedings{Registration-Arikan,
   author={M. Arikan and A. Sadeghipour and B. Gerendas and R. Told and U. Schmidt-Erfurt},
   year={2019},
   title={Deep learning based multi-modal registration for retinal imaging},
   booktitle={Interpretability of Machine Intelligence in Medical Image Computing and Multimodal Learning for Clinical Decision Support},
   pages={75-82},
}

@inproceedings{Registration-Tian,
   author={Y. Tian et al.},
   year={2020},
   title={Multi-scale U-Net with edge guidance for multimodal retinal image deformable registration},
   booktitle={IEEE Eng. Med. Biol. Soc. (EMBC)},
   pages={1360–1363},
}

@inproceedings{Registration-Lee,
   author={J. Lee and P. Liu and J. Cheng and H. Fu},
   year={2019},
   title={A deep step pattern representation for multimodal retinal image registration},
   booktitle={IEEE/CVF Int. Conf. Comput. Vis. (ICCV)},
   pages={5076-5085},
}

@inproceedings{Registration-SuperJunction,
   author={Y. Wang and X. Wang and Z. Gu and W. Liu and W. Ng and W. Huang and J. Cheng},
   year={2024},
   title={Superjunction: Learning-based junction detection for retinal image registration.},
   booktitle={AAAI Conference on Artificial Intelligence},
   pages={292–300},
}

@article{Registration-Liang-EyeKey,
  author  = {Y. Liang and D. Ma and X. Wu},
  title   = {{EyeKey}: Self-Supervised Keypoint Detection and Description Network Based on Local Feature Saliency for Retinal Image Global Registration},
  journal = {IEEE Transactions on Image Processing},
  volume  = {35},
  pages   = {4772--4787},
  year    = {2026},
  doi     = {10.1109/TIP.2026.3688165}
}

@inproceedings{Registration-Liu-GAMorph,
  author    = {Y. Liu and B. Yu and T. Chen and Y. Gu and B. Du and Y. Xu and J. Cheng},
  title     = {Progressive Retinal Image Registration via Global and Local Deformable Transformations},
  booktitle = {2024 IEEE International Conference on Bioinformatics and Biomedicine (BIBM)},
  pages     = {2183--2190},
  year      = {2024},
  doi       = {10.1109/BIBM62325.2024.10821896}
}

@INPROCEEDINGS{Registration-Wen-ICIP,
  author={J. Zhang and B. Wen and F. Kalaw and M. Cavichini and D. Bartsch and W. Freeman and T. Nguyen and C. An},
  booktitle={2023 IEEE International Conference on Image Processing (ICIP)}, 
  title={Accurate Registration between Ultra-Wide-Field and Narrow Angle Retina Images with 3D Eyeball Shape Optimization}, 
  year={2023},
  pages={2750-2754},
}

@inproceedings{Registration-CrossRAFT,
  author    = {S. Zhou and W. Tan and B. Yan},
  title     = {Promoting Single-Modal Optical Flow Network for Diverse Cross-Modal Flow Estimation},
  booktitle = {Proceedings of the AAAI Conference on Artificial Intelligence},
  volume    = {36},
  number    = {3},
  pages     = {3562--3570},
  year      = {2022},
  doi       = {10.1609/aaai.v36i3.20268}
}

@inproceedings{Registration-MCMRAFT,
  author    = {M. Zhai and K. Ni and J. Xie and H. Gao},
  title     = {Cross-Modal Optical Flow Estimation via Modality Compensation and Alignment},
  booktitle = {2023 IEEE International Conference on Acoustics, Speech and Signal Processing (ICASSP)},
  pages     = {1--5},
  year      = {2023},
  doi       = {10.1109/ICASSP49357.2023.10095898}
}

@article{Registration-Zhang-JSTARS,
  author  = {H. Zhang and L. Lei and W. Ni and X. Yang and T. Tang
             and K. Cheng and D. Xiang and G. Kuang},
  title   = {Optical and {SAR} Image Dense Registration Using a Robust Deep Optical Flow Framework},
  journal = {IEEE Journal of Selected Topics in Applied Earth Observations and Remote Sensing},
  volume  = {16},
  pages   = {1269--1294},
  year    = {2023},
  doi     = {10.1109/JSTARS.2023.3235535}
}

@article{Registration-Sun-GDROS,
  author  = {Z. Sun and S. Zhi and R. Li and J. Xia and Y. Liu and W. Jiang},
  title   = {{GDROS}: A Geometry-Guided Dense Registration Framework for Optical--{SAR} Images Under Large Geometric Transformations},
  journal = {IEEE Transactions on Geoscience and Remote Sensing},
  volume  = {63},
  pages   = {1--15},
  year    = {2025},
  doi     = {10.1109/TGRS.2025.3627132}
}

@inproceedings{Registration-Zhang-DCFlow,
  author    = {R. Zhang and J. Wang and S.-Y. Cao and Z. Yu and J. Yu
               and G. Zhang and H.-L. Shen},
  title     = {Rethinking Unsupervised Cross-Modal Flow Estimation: Learning from Decoupled Optimization and Consistency Constraint},
  booktitle = {International Conference on Learning Representations},
  year      = {2026}
}

@inproceedings{Registration-Liu-CRFT,
  author    = {X. Liu and M. Ding and Z. Sun and Z. Li and X. Teng},
  title     = {{CRFT}: Consistent-Recurrent Feature Flow Transformer for Cross-Modal Image Registration},
  booktitle = {Proceedings of the IEEE/CVF Conference on Computer Vision and Pattern Recognition},
  pages     = {34784--34794},
  year      = {2026}
}

@article{Registration-Sivaraman-RetRegNet,
        title = {RetinaRegNet: A zero-shot approach for retinal image registration},
        journal = {Computers in Biology and Medicine},
        volume = {186},
        pages = {109645},
        year = {2025},
        author = {V. Sivaraman and M. Imran and Q. Wei and P. Muralidharan and M. Tamplin and I. Grumbach and R. Kardon and J. Wang and Y. Zhou and W. Shao},
}

@inproceedings{Registration-SuperPoint,
   author={D. DeTone and T. Malisiewicz and A. Rabinovich},
   year={2018},
   title={SuperPoint: Self-supervised interest point detection and description},
   booktitle={IEEE/CVF Conf. Comput. Vis. Pattern Recognit. Workshops (CVPRW)},
   pages={224-236},
}

@article{Registration-Klein-TIP,
  author  = {S. Klein and M. Staring and J. P. W. Pluim},
  title   = {Evaluation of Optimization Methods for Nonrigid Medical Image Registration Using Mutual Information and {B}-Splines},
  journal = {IEEE Transactions on Image Processing},
  volume  = {16},
  number  = {12},
  pages   = {2879--2890},
  year    = {2007},
  doi     = {10.1109/TIP.2007.909412}
}

@article{Registration-Wang-Metrics,
  author  = {Y. Wang and J. Zhang and M. Cavichini and D.-U. G. Bartsch and W. R. Freeman and T. Q. Nguyen and C. An},
  title   = {Study on Correlation Between Subjective and Objective Metrics for Multimodal Retinal Image Registration},
  journal = {IEEE Access},
  volume  = {8},
  pages   = {190897--190905},
  year    = {2020},
  doi     = {10.1109/ACCESS.2020.3032348}
}

@article{Registration-Kalaw-Eye,
  author  = {F. G. P. Kalaw and M. Cavichini and J. Zhang and B. Wen and A. C. Lin and A. Heinke and T. Nguyen and C. An and D.-U. G. Bartsch and L. Cheng and W. R. Freeman},
  title   = {Ultra-Wide Field and New Wide Field Composite Retinal Image Registration with {AI}-Enabled Pipeline and {3D} Distortion Correction Algorithm},
  journal = {Eye},
  volume  = {38},
  number  = {6},
  pages   = {1189--1195},
  year    = {2024},
}

@article{Registration-Klein-Elastix,
  author  = {S. Klein and M. Staring and K. Murphy and M. A. Viergever and J. P. W. Pluim},
  title   = {{elastix}: A Toolbox for Intensity-Based Medical Image Registration},
  journal = {IEEE Transactions on Medical Imaging},
  volume  = {29},
  number  = {1},
  pages   = {196--205},
  year    = {2010},
  doi     = {10.1109/TMI.2009.2035616}
}

@inproceedings{Registration-STN,
   author={M. Jaderberg and K. Simonyan and A. Zisserman},
   year={2015},
   title={Spatial transformer networks},
   booktitle={2015 Advances in Neural Information Processing Systems (NIPS)},
   pages={2017-2025},
}

@phdthesis{Registration-Song-Thesis,
  author  = {Joo Hyun Song},
  title   = {Methods for Evaluating Image Registration},
  school  = {University of Iowa},
  address = {Iowa City, IA, USA},
  year    = {2017},
  doi     = {10.17077/etd.v0vailob}
}

@inproceedings{Correspondence-Teed-RAFT,
  author    = {Z. Teed and J. Deng},
  title     = {{RAFT}: Recurrent All-Pairs Field Transforms for Optical Flow},
  booktitle = {European Conference on Computer Vision},
  pages     = {402--419},
  year      = {2020},
}

@inproceedings{Correspondence_CLe_CVPR,
   author={K. M. Yi and E. Trulls and Y. Ono and V. Lepetit and M. Salzmann and P. Fua},
   year={2018},
   title={Learning to find good correspondences},
   booktitle={IEEE/CVF Conf. Comput. Vis. Pattern Recognit.},
   pages={2666–2674},
}

@ARTICLE{Correspondence_MIFNet_TIP,
  author={Y. Liu and Z. Sun and B. Yu and Y. Zhao and B. Du and Y. Xu and J. Cheng},
  journal={IEEE Transactions on Image Processing}, 
  title={MIFNet: Learning Modality-Invariant Features for Generalizable Multimodal Image Matching}, 
  year={2025},
  volume={34},
  pages={3593-3608},
}

@inproceedings{Correspondence-Ren-MINIMA,
  author    = {J. Ren and X. Jiang and Z. Li and D. Liang and X. Zhou and X. Bai},
  title     = {{MINIMA}: Modality Invariant Image Matching},
  booktitle = {Proceedings of the IEEE/CVF Conference on Computer Vision and Pattern Recognition},
  pages     = {23059--23068},
  year      = {2025}
}

@inproceedings{Correspondence-Amir-ECCVW,
  author    = {S. Amir and Y. Gandelsman and S. Bagon and T. Dekel},
  title     = {On the Effectiveness of {ViT} Features as Local Semantic Descriptors},
  booktitle = {Computer Vision -- ECCV 2022 Workshops},
  pages     = {39--55},
  year      = {2023},
  doi       = {10.1007/978-3-031-25069-9_3}
}

@inproceedings{Correspondence-Zhang-NeurIPS,
  author    = {J. Zhang and C. Herrmann and J. Hur and L. Polania Cabrera and V. Jampani and D. Sun and M.-H. Yang},
  title     = {A Tale of Two Features: Stable Diffusion Complements {DINO} for Zero-Shot Semantic Correspondence},
  booktitle = {Advances in Neural Information Processing Systems},
  volume    = {36},
  year      = {2023}
}

@inproceedings{Correspondence-DIFT-NeurIPS,
  author    = {L. Tang and M. Jia and Q. Wang and C. P. Phoo and B. Hariharan},
  title     = {Emergent Correspondence from Image Diffusion},
  booktitle = {Advances in Neural Information Processing Systems},
  volume    = {36},
  pages     = {1363--1389},
  year      = {2023}
}

@inproceedings{Correspondence-SuperGlue-CVPR,
   author={P. Sarlin and D. DeTone and T. Malisiewicz and A. Rabinovich},
   year={2020},
   title={Superglue: Learning feature matching with graph neural networks},
   booktitle={IEEE/CVF conference on computer vision and pattern recognition (CVPR)},
   pages={4938-4947},
}

@inproceedings{Correspondence-PWCNet-CVPR,
  title     = {{PWC-Net}: {CNNs} for Optical Flow Using Pyramid, Warping, and Cost Volume},
  author    = {D. Sun and X. Yang and M. Liu and J. Kautz},
  booktitle = {Proceedings of the IEEE Conference on Computer Vision and Pattern Recognition (CVPR)},
  pages     = {8934--8943},
  year      = {2018}
}

@ARTICLE{Segmentation_Wen_TIP,
  author={B. Wen and A. Heinke and A. Agnihotri and D. Bartsch and W. Freeman and T. Nguyen and C. An},
  journal={IEEE Transactions on Image Processing}, 
  title={Universal Vessel Segmentation for Multi-Modality Retinal Images}, 
  year={2025},
  volume={34},
  pages={7903-7918},
}

@misc{Application-eye_tracking,
 author = {B. Wen and D. Lohr and Y. An and P. Anand and A. Fix and R. Qian and C. Fromm and Y. Ding and T. Nguyen and M. El-Haddad and F. La Rocca},
 title = {Establishing Robust Retinal Eye Tracking: A Weakly Supervised Algorithmic Framework},
 note = {arXiv preprint arXiv:2605.09181},
 year = {2026}
}

@article{Application-stability,
author = {Q. Yang and J. Zhang and K. Nozato and K. Saito and D. Williams and A. Roorda and E. Rossi},
journal = {Biomed. Opt. Express},
number = {9},
pages = {3174--3191},
publisher = {Optica Publishing Group},
title = {Closed-loop optical stabilization and digital image registration in adaptive optics scanning light ophthalmoscopy},
volume = {5},
year = {2014}
}

@article{Dataset-FIRE,
  author  = {C. Hernandez-Matas and X. Zabulis and A. Triantafyllou and P. Anyfanti and S. Douma and A. A. Argyros},
  title   = {{FIRE}: Fundus Image Registration Dataset},
  journal = {Modeling and Artificial Intelligence in Ophthalmology},
  volume  = {1},
  number  = {4},
  pages   = {16--28},
  year    = {2017},
}

@article{Dataset-CFFA,
author = {S. H. M. Alipour and H. Rabbani and M. R. Akhlaghi},
title = {Diabetic retinopathy grading by digital curvelet transform},
journal = {Comput. Math. Methods Med.},
volume = 2012,
number = 1,
pages = {1607–1614},
year = 2016
}

@inproceedings{Others-stable_diffusion,
  author    = {R. Rombach and A. Blattmann and D. Lorenz and P. Esser
               and B. Ommer},
  title     = {High-Resolution Image Synthesis with Latent Diffusion Models},
  booktitle = {Proceedings of the IEEE/CVF Conference on Computer Vision
               and Pattern Recognition},
  pages     = {10684--10695},
  year      = {2022},
}

@article{Others-DINOV3,
  author  = {O. Sim{\'e}oni and H. V. Vo and M. Seitzer and F. Baldassarre
             and M. Oquab and C. Jose and V. Khalidov and M. Szafraniec
             and S. Yi and M. Ramamonjisoa and F. Massa and D. Haziza
             and L. Wehrstedt and J. Wang and T. Darcet and T. Moutakanni
             and L. Sentana and C. Roberts and A. Vedaldi and J. Tolan
             and J. Brandt and C. Couprie and J. Mairal and H. J{\'e}gou
             and P. Labatut and P. Bojanowski},
  title   = {{DINOv3}},
  journal = {Transactions on Machine Learning Research},
  year    = {2026}
}

@inproceedings{Others-CycleGAN,
author = {J. Zhu and T. Park and P. Isola and A. Efros},
title = {Unpaired image-to-image translation using cycle-consistent adversarial networks},
booktitle = {IEEE/CVF International Conference on Computer Vision},
pages = {2223--2232},
year = {2017}
}

@ARTICLE{Others_ResdoU-Net,
  title={ResDO-UNet: A deep residual network for accurate retinal vessel segmentation from fundus images},
  author={Y. Liua and J. Shena and L. Yanga and G. Biana and H. Yu},
  journal={Biomedical Signal Processing and Control},
  year={2023},
  volume={79},
  pages={104087},
}

\vfill

\end{document}